\documentclass[%
 aip,
 amsmath,amssymb,
reprint,%
]{revtex4-1}

\usepackage{graphicx}
\usepackage{dcolumn}
\usepackage{bm}

\usepackage[utf8]{inputenc}
\usepackage[T1]{fontenc}
\usepackage{mathptmx}
\usepackage{etoolbox}

\usepackage{amsmath}
\usepackage{amssymb}
\usepackage{mathrsfs}
\usepackage{mathtools}

\usepackage{csquotes}
\usepackage{hyperref}

\usepackage{multirow}

\newcommand{\tensg}[1]{\boldsymbol{\underline{#1}}}

\makeatletter
\def\@email#1#2{%
 \endgroup
 \patchcmd{\titleblock@produce}
  {\frontmatter@RRAPformat}
  {\frontmatter@RRAPformat{\produce@RRAP{*#1\href{mailto:#2}{#2}}}\frontmatter@RRAPformat}
  {}{}
}%
\makeatother
\begin{document}

\preprint{AIP/123-QED}

\title[Nonreciprocal pattern-forming systems with O(2)-symmetry]{Higher-codimension points as organizing centers in nonreciprocal pattern-forming systems with O(2)-symmetry}

\author{Yuta Tateyama}
\thanks{These authors contributed equally to this work.}
\affiliation{Department of Physics, Chiba University, Chiba 263-8522, Japan}

\author{Daniel Greve}
\thanks{These authors contributed equally to this work.}
\affiliation{Institute of Theoretical Physics, University of M\"unster, Wilhelm-Klemm-Strasse 9, 48149 M\"unster, Germany}

\author{Hiroaki Ito}
\affiliation{Department of Physics, Chiba University, Chiba 263-8522, Japan}

\author{Shigeyuki Komura}
\affiliation{Zhejiang Key Laboratory of Soft Matter Biomedical Materials, Wenzhou Institute, University of Chinese Academy of Sciences, Wenzhou, Zhejiang 325000, China}

\author{Hiroyuki Kitahata}
\affiliation{Department of Physics, Chiba University, Chiba 263-8522, Japan}

\author{Uwe Thiele}
\email{u.thiele@uni-muenster.de}
\homepage{http://www.uwethiele.de}
\affiliation{Institute of Theoretical Physics, University of M\"unster, Wilhelm-Klemm-Strasse 9, 48149 M\"unster, Germany}
\affiliation{Center for Data Science and Complexity (CDSC), University of M\"unster, Corrensstrasse 2, 48149 M\"unster, Germany}

\date{\today}

\begin{abstract}
Focusing on a two-field Swift-Hohenberg model with linear nonreciprocal interactions, this study investigates how emerging higher-codimension points act as organizing centers for the nonequilibrium phase diagram that features various steady and dynamic phases.
Complementing the numerical analysis of the field equations with time simulations and path continuation techniques, we derive a reduced dynamical system corresponding to a one-mode approximation for the critical-wavenumber modes.
Furthermore, we derive the normal form equations that are valid in the vicinity of the Takens-Bogdanov bifurcation with O(2)-symmetry, which allows us to draw on corresponding literature results.
Comparing results obtained on the different levels of description, we discuss the bifurcation structure relating trivial uniform and inhomogeneous steady states as well as traveling, standing and modulated waves.
We also contextualize the relevance of recently highlighted features of the linear mode structure, i.e., of the dispersion relations, termed \enquote{critical exceptional points} for the transitions between the nonequilibrium phases.
\end{abstract}

\maketitle

\begin{quotation}
    Newton's third law states that for every action, there is an equal and opposite reaction.
    Nevertheless, in the world of \enquote{active matter,} for example, bacterial swarms and chemical mixtures, this rule often seems to crumble.
    Envision a predator species chasing prey that never reciprocates the pursuit.
    While such one-way interactions trigger rich and complicated behaviors, such as traveling bands or oscillatory waves, the mathematical framework needed to describe them is deeply rooted in the classical bifurcation theory of systems with spatial symmetries.
    This study employs the nonreciprocal Swift-Hohenberg model as a representative nonreciprocal pattern-forming system to bridge the gap to established results of advanced nonlinear dynamics.
    Our analysis reveals that, mathematically, the Takens-Bogdanov bifurcation with O(2)-symmetry is the most important organizing center for the route from static stripes to traveling waves.
\end{quotation}


\section{Introduction}
In recent years, nonreciprocity in the relation of different species has emerged as a unifying and generative concept in the study of collective behavior and pattern formation in multispecies systems that are permanently out of equilibrium.
Thus, \enquote{nonreciprocity} often refers to an effective breakdown of Newton's third law, the symmetry of action and reaction.
This not only signals a departure from the conservative, time-reversible dynamics but also, for an overdamped system, a departure from a relaxational dynamics that ultimately approaches a steady equilibrium state.
Besides this mechanical nonreciprocity, we distinguish thermodynamic nonreciprocity---which violates Onsager's reciprocity relations between thermodynamic mobilities---and chemical nonreciprocity---which violates detailed balance in chemical reactions (thereby violating de Donder's relations).
All of these terms refer to effective interspecies relations.

Common microscopic and mesoscopic mechanisms that give rise to effective nonreciprocal interactions between species include nonreciprocal pairwise forces between particles~\cite{IBHD2015prx}, quorum-sensing in multispecies bacterial colonies~\cite{PeGG2016bmb,DOCT2023nc,DAGM2023prl}, nonreciprocal alignment of rodlike active particles~\cite{KrKl2022njp} or mediated interactions between consumers and resources~\cite{BlRM2024prl}.
Guided by this emerging focus, a common approach to such active systems is the formulation of two-species nonreciprocal field theories with a simple linear nonreciprocal coupling to capture the relevant qualitative behavior while retaining a meaningful connection to the passive (\enquote{dead}) limit, where the coupling is reciprocal and conventional thermodynamic intuition applies.
Prominent examples of active multicomponent field theories are nonreciprocal Allen-Cahn~\cite{LHKH2023jpsj,SeMM2021nc}, Cahn-Hilliard~\cite{SaAG2020prx,YoBM2020pnasusa,FrWT2021pre,BrMa2024prx}, Swift-Hohenberg~\cite{SATB2014c,BFNR2018c,FHLV2021n,FHKG2023pre,TIKK2024pre,RHST2025arxiv} and phase-field-crystal (PFC) models~\cite{HAGK2021ijam}.
Note that models of different types can also be combined into further classes of nonreciprocal models, e.g., the nonreciprocal coupling of a Cahn-Hilliard and a Swift-Hohenberg equation~\cite{FHKG2023pre}, and may, in principle, all be derived from microscopic models, e.g., from nonreciprocal Ising models~\cite{avni2025dynamical,avni2025nonreciprocal,BlTG2026sp}.
Linear interactions often facilitate further (semi-)analytic progress\cite{BrMa2024prx,GLFT2025prl,FHKG2023pre}.
In particular, they may result in a \enquote{spurious gradient dynamics structure} which can be used to, e.g., derive conditions for phase coexistence~\cite{GLFT2025prl}, and support features such as resting asymmetric states that are not generic for nonvariational models~\cite{FHKG2023pre}.

A key object of such studies is the occurrence of dynamic phases that are directly caused by nonreciprocity, as well as of static phases that do not occur in the corresponding reciprocal model.
An example is the suppression of coarsening in nonreciprocal Cahn-Hilliard models~\cite{FrWT2021pre} that is directly related to the existence of a conserved-Turing instability, localized states and crystalline phases~\cite{FrTh2021ijam,GLFT2025prl}.
The dynamic phases are related to the widely studied onset of oscillatory behavior in the form of regular traveling, standing, and modulated waves or chaotic oscillations~\cite{LHKH2023jpsj,SaAG2020prx,YoBM2020pnasusa,FrWT2021pre,BrMa2024prx,SATB2014c,FHLV2021n,FHKG2023pre,TIKK2024pre,GLFT2025prl}.
More abstractly, for traveling waves this is sometimes discussed as the spontaneous breaking of parity-time (PT) symmetry~\cite{FHLV2021n,SuKL2023pre,BrMa2024prx}.
A common property of the Swift-Hohenberg models and their conserved relatives, the PFC models, is the possibility of a nonzero wavenumber at instability onset in the respective passive (reciprocal) limiting case~\cite{CrHo1993rmp,TARG2013pre}, i.e., of Turing and conserved-Turing instabilities, respectively.
In the nonreciprocal (and other nonvariational) variants these instabilities can become oscillatory,  i.e., they correspond to wave and conserved-wave instabilities, respectively~\footnote{For the classification and nomenclature of instabilities used here, see the supplement of Ref.~\onlinecite{FrTh2023prl}.}. The resulting steady patterns, traveling waves, modulated waves, and standing waves can all be described by coupled amplitude equations obtained in a one-mode approximation. The resulting dynamical system (ordinary differential equations) inherits an O(2)-symmetry from the isotropy and homogeneity of space underlying the field equations.

Interestingly, the theoretical study of pattern-forming instabilities and their relations in out-of-equilibrium systems with continuous symmetries, in particular an O(2)-symmetry (related to translation), has a rich history rooted in fluid dynamics and reaction-diffusion systems~\cite{CrHo1993rmp, golubitsky1984symmetries, sasa1990model, yang2002pattern}.
Dating back several decades before the present discussions of explicitly nonreciprocal interactions, a robust analytical machinery was developed to understand bifurcations and mode interactions in such systems~\cite{GolubitskyStewart2002}.
In particular, this applies to local and global codimension-one bifurcations related to transitions to the above mentioned dynamic phases, e.g., Hopf, drift-pitchfork, homoclinic, and saddle-node infinite period (SNIPer) bifurcations, as well as to the codimension-two bifurcations which act as organizing centers of the transition between stationary and oscillatory pattern formation, e.g., the Takens-Bogdanov (or double-zero) bifurcation~\cite{takens1974singularities,bogdanov1975versal,Wiggins2003,Kuzn2005ijbc,Kuznetsov2023}.

The weakly nonlinear behavior in the vicinity of a Hopf bifurcation with O(2)-symmetry was fully classified in Refs.~\onlinecite{GoRo1987jde,CrKn1988pd}.
The mechanism of the drift-pitchfork bifurcation, i.e., the spontaneous onset of motion when an additional damped mode becomes neutral and exactly matches the Goldstone mode of translation or rotation, was identified and analyzed~\cite{CLHL1990prl,GGGC1991pra,FaDT1991jpi,KnMo1990pra,KrMi1994prl,MoLP2003el,Liehr2013}~\footnote{In a competing terminology that recently gained some traction driven by notions diffusing in from the quantum world, this is now sometimes classified as an \enquote{exceptional transition} related to a \enquote{critical exceptional point\cite{FHLV2021n,SuKL2023pre, SuKL2023prl}}.
This reflects that a related transition from real to complex eigenvalues is also discussed in non-Hermitian quantum mechanics and nonlinear optics~\cite{MiAl2019s,AsGU2020ap,MeAL2024apl}.}.
Note that the drift-pitchfork bifurcation is also sometimes simply called a \enquote{traveling bifurcation}, e.g., in the context of the Ising-Bloch transition for fronts in reaction-diffusion systems and nonlinear optics~\cite{HaMe1994c,MPLS2001pre,Liehr2013,Pismen2023}.
Even for the Takens-Bogdanov bifurcation in systems with O(2)-symmetry, crucial for the present work, all cases without nonlinear degeneracy were classified~\cite{DaKn1987ptrslsapes, RuKn2017ds}.

At first glance, it might seem odd that readily available universal results are rarely seen directly applied or discussed in detail in the context of \enquote{nonreciprocal phase transitions}~\cite{FHLV2021n,KrKl2022njp,KrKl2024prl,KrKl2025cp,BrMa2024prx,YoBM2020pnasusa,SuKL2023prl} or, indeed, other specific recent contexts.
We attribute this to the fact that the pursuit of universality comes at a price: Analytical progress is often achieved either by the heavy use of nonlinear near-identity transformations of a center manifold and the introduction of unfolding parameters in a later stage of the analysis or by starting immediately from a normal form postulated based on symmetry.
First, this entails the technical complication that the connection between the unfolding parameters and the parameters of the original physical microscopic discrete or macroscopic continuous model is often rather indirect and must be laboriously traced through multistep calculations and transformations.
Second, strictly speaking, the universal analytic results are only valid in the vicinity of specific points in parameter space, such that an overview of its entirety is rarely achieved.
Third, an inherent weakness of these approaches comes with global bifurcations, which can often not be captured by a local phase-space analysis.

In the present work, we apply and discuss these universal results in the context of systems with a dominant characteristic wavelength and nonreciprocal interactions.
Thereby, we revisit the two-field Swift-Hohenberg model with linear nonreciprocal interactions~\cite{SATB2014c, BFNR2018c, FHLV2021n, FHKG2023pre, TIKK2024pre}.
In particular, we consider its phase behavior on three levels: (i) the original partial differential equations, (ii) an ordinary differential equation model obtained via a one-mode approximation, and (iii) normal forms valid in the vicinity of the organizing codimension-two bifurcations.
The latter are obtained from (ii) via near-identity transformations, and allow for connection to and drawing on literature results~\cite{DaKn1987ptrslsapes}.
We then compare and relate the results obtained on the different levels of description with a focus on the \enquote{routes to traveling states}\cite{YoBM2020pnasusa}, i.e., on the rich bifurcation structure responsible for the nonreciprocity-induced transitions from inhomogeneous steady states to traveling waves, which can include standing and modulated waves at intermediate stages.
In passing we will emphasize that drift-pitchfork bifurcations (related to critical exceptional points of codimension one) are just one of potentially many bifurcations involved, but that Takens-Bogdanov bifurcations with O(2)-symmetry (related to a critical exceptional point of codimension two) indeed organize the entire transition at the onset of pattern formation.

Our work is structured as follows: In Section~\ref{sec:gov_eqs} we present the nonreciprocal Swift-Hohenberg model and the dynamical system that is obtained as a one-mode approximation.
The subsequent Section~\ref{sec:phase} presents a fixed point and bifurcation analysis of both models as well as selected bifurcation, stability and phase diagrams.
This clarifies the role of the individual steady and dynamic states in the overall transition.
Then, Section~\ref{sec:TB} focuses on the Takens-Bogdanov bifurcation, derives the corresponding normal form, and discusses its relevance as an organizing center for pattern formation in nonreciprocal systems.
Section~\ref{sec:param_space} structures fully nonlinear regimes of our model in terms of further bifurcations of higher codimension, which ultimately allows one to classify all different routes to traveling states.
Finally, the discussion in Section~\ref{sec:conc} concludes the work.

\section{Governing equations}\label{sec:gov_eqs}
\begin{figure*}[ht]
    \centering
    \includegraphics[width=\textwidth]{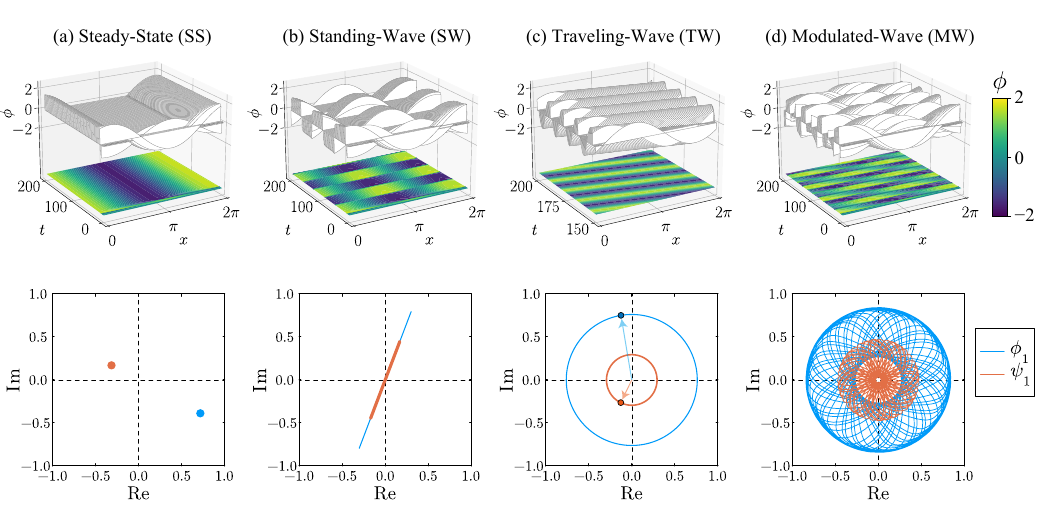}
    \caption{
        Simulation results of the (top row) one-dimensional NRSH model~\eqref{eq:NRSH} as spacetime plots of the order parameter field $\phi(x,t)$ and (bottom row) as corresponding trajectories in the complex plane of the amplitudes $\phi_1(t)$ and $\psi_1(t)$ , both at $(\varepsilon, \chi, \delta)=(1, 1, 0)$.
        (a) Steady-State (SS) at $\alpha = 1.27$, (b) Standing-Wave (SW) at $\alpha = 1.28$, (c) Traveling-Wave (TW) at $\alpha = 1.35$, and (d) Modulated-Wave (MW) at $\alpha = 1.29$.
    }
    \label{fig:sptm-PDE-trajectories-ODE}
\end{figure*}

The one-dimensional nonreciprocal Swift-Hohenberg (NRSH) model, which describes the dynamics of two real order parameter fields $\phi(x,t)$ and $\psi(x,t)$ coupled through reciprocal and nonreciprocal interactions, is given by~\cite{SATB2014c,BFNR2018c,FHLV2021n,FHKG2023pre,TIKK2024pre,RHST2025arxiv}:
\begin{subequations}
    \begin{align}
        \partial_t \phi &= [\varepsilon + \delta - (k_{\mathrm{c}}^2 + \partial_x^2)^2] \phi - \phi^3 - (\chi + \alpha) \psi,
        \label{eq:NRSH-phi}
        \\
        \partial_t \psi &= [\varepsilon - \delta - (k_{\mathrm{c}}^2 + \partial_x^2)^2] \psi - \psi^3 - (\chi - \alpha) \phi,
        \label{eq:NRSH-psi}
    \end{align}\label{eq:NRSH}
\end{subequations}
where $\partial_t$ and $\partial_x$ denote the partial time and space derivatives, respectively.
Here, we consider a one-dimensional system with periodic boundary conditions.
The parameters $\varepsilon$ and $\delta$ denote the average potential depth and the difference in potential depth relative to the conventional Swift-Hohenberg model; the local free energy is given by $\phi^4 / 4 - (\varepsilon + \delta) \phi^2 / 2 + \psi^4 / 4 - (\varepsilon - \delta) \psi^2 / 2$.
The parameters $\chi$ and $\alpha$ correspond to the reciprocal and nonreciprocal coupling strengths, respectively.
In comparison to Ref.~\onlinecite{TIKK2024pre}, we extend the model by introducing the parameter $\delta$, such that $\varepsilon$, $\delta$, $\chi$ and $\alpha$ characterize the entries of a general $2 \times 2$ linear interaction matrix.

Since both species in the NRSH model~\eqref{eq:NRSH} are assumed to have the same critical wavenumber $k_{\mathrm{c}}$, the emerging spatiotemporal patterns are well approximated by the amplitudes of the critical spatial mode.
Thus, we retain only the complex amplitudes $\phi_1$ and $\psi_1$ of the primary spatial mode for the real order parameters $\phi$ and $\psi$, i.e., $\phi(x,t) \simeq \phi_1(t) e^{- \mathrm{i} k_{\mathrm{c}} x} + \mathrm{c.c.}$ and $\psi(x,t) \simeq \psi_1(t) e^{- \mathrm{i} k_{\mathrm{c}} x} + \mathrm{c.c.}$.
Using this one-mode approximation, we obtain the following reduced system of two coupled ordinary differential equations~\cite{TIKK2024pre}
\begin{subequations}\label{eq:nr_AE}
    \begin{align}
        \dot{\phi}_1 &= (\varepsilon + \delta) \phi_1 - (\chi + \alpha) \psi_1 - 3 |\phi_1|^2 \phi_1,
        \label{eq:phi_1}
        \\
        \dot{\psi}_1 &= (\varepsilon - \delta) \psi_1 - (\chi - \alpha) \phi_1 - 3 |\psi_1|^2 \psi_1.
        \label{eq:psi_1}
    \end{align}
\end{subequations}
Here the overdot indicates the time derivative.
On the amplitude level, the spatial translation symmetry $x \mapsto x + a$ of the NRSH model \eqref{eq:NRSH} corresponds to a symmetry under simultaneous phase shifts, i.e., the one-mode approximation~\eqref{eq:nr_AE} is invariant under $(\phi_1, \psi_1) \mapsto (e^{\mathrm{i} \Theta} \phi_1, e^{\mathrm{i} \Theta} \psi_1)$ with $\Theta \in [0,2\pi)$.
Similarly, spatial parity symmetry $x \mapsto -x$ corresponds to symmetry under simultaneous complex conjugation $(\phi_1,\psi_1)\mapsto (\bar{\phi}_1,\bar{\psi}_1)$.
On both levels of description, the combination of the two symmetries corresponds to the semidirect product $\mathrm{O}(2) \cong \mathrm{U}(1) \rtimes \mathbb{Z}_2$.

Apart from the time-independent Null (N) phase $(\phi,\psi)=(0,0)$ that inherits the full spatial O(2)-symmetry, we observe four different phases with spontaneously broken continuous translation symmetry.
Their characteristic patterns on both levels of description, i.e., in the full spatiotemporal description via Eq.~\eqref{eq:NRSH} and in the one-mode approximation~\eqref{eq:nr_AE} are shown in the upper and lower row of Fig.~\ref{fig:sptm-PDE-trajectories-ODE}, respectively (see Appendix~\ref{app:direct_sim} for details of the employed numerical approaches).

Figure~\ref{fig:sptm-PDE-trajectories-ODE}(a) shows the Steady-State (SS) phase, characterized by a spatially periodic static pattern. The steady state is time-invariant and retains the parity symmetry of the underlying system.
Figure~\ref{fig:sptm-PDE-trajectories-ODE}(b) shows the Standing-Wave (SW) phase, which is no longer time-invariant but also maintains the parity symmetry.
In contrast, traveling-waves (TW) shown in Fig.~\ref{fig:sptm-PDE-trajectories-ODE}(c) break both time-invariance and spatial symmetries.
Left- and right-traveling waves are related under the combined inversion of space and time.
The Modulated-Wave (MW) phase is characterized by a spatially propagating wave with an oscillating amplitude as shown in Fig.~\ref{fig:sptm-PDE-trajectories-ODE}(d), and retains no symmetries.
Note that in the context of active matter systems, SW, TW, and MW phases are often referred to as \textit{swap}, \textit{chiral}, and \textit{chiral-swap} phases, respectively\cite{FHLV2021n}.

Since the spontaneous breaking of the continuous spatial translation symmetry results in a Goldstone zero mode, Eqs.~\eqref{eq:nr_AE} can be reduced from four to three effective degrees of freedom by setting $\phi_1 = \rho_\phi e^{\mathrm{i} \theta_\phi}$ and $\psi_1 = \rho_\psi e^{\mathrm{i} \theta_\psi}$, and introducing the phase difference $\eta = \theta_\psi - \theta_\phi$. The fourth degree of freedom then corresponds to motion along the group orbit of the continuous symmetry, i.e., a spatial shift of the pattern on the level of the NRSH equations \eqref{eq:NRSH} and a simultaneous rotation of both amplitudes in the complex plane in the one-mode approximation \eqref{eq:nr_AE}.
The remaining system of ordinary differential equations for the real dynamic variables $\rho_\phi, \rho_\psi$  and $\eta$ is given by:
\begin{subequations}\label{eq:phase_difference}
\begin{align}
    \dot{\rho}_{\phi} &= (\varepsilon + \delta - 3 \rho_\phi^2
     ) \rho_\phi - (\chi + \alpha) \rho_\psi \cos \eta,
    \\
    \dot{\rho}_\psi &= (\varepsilon - \delta - 3 \rho_\psi^2
    ) \rho_\psi - (\chi - \alpha) \rho_\phi \cos \eta,
    \\
    \dot{\eta} &= \left[
        (\alpha + \chi) \frac{\rho_\psi}{\rho_\phi} - (\alpha - \chi) \frac{\rho_\phi}{\rho_\psi}
    \right] \sin \eta.\label{eq:phase_ev}
\end{align}
\end{subequations}
The dynamics of $\theta_\phi$ and $\theta_\psi$ are fully determined by Eqs.~\eqref{eq:phase_difference} and given by $\dot{\theta}_\phi = -(\chi + \alpha) (\rho_\psi /\rho_\phi) \sin \eta$ and $\dot{\theta}_\psi = (\chi - \alpha) (\rho_\phi / \rho_\psi) \sin \eta$, respectively.

\begin{table}[htbp]
    \centering
    \caption{
        Bifurcation conditions obtained by the bifurcation analyses of the one-mode approximation~\eqref{eq:nr_AE} and from the normal form of the Takens-Bogdanov bifurcation with O(2)-symmetry~\eqref{eq:TB-normal} discussed in Section~\ref{sec:TB}.
        The derivations and resulting equations are given in Appendices~\ref{app:Turing-wave} and \ref{app:local-bif-derivation}.
    }
    \begin{tabular}{c||c|c|c}
        Bifurcation & Related phases & One-mode & TB
        \\
        \hline
        Turing (T) & N \& SS & Eq.~\eqref{eq:Turing-bifurcation}&
        $\mu=0$\\[.3em]
        wave (W) & N \& TW/SW & Eq.~\eqref{eq:wave-bifurcation}&
        $\nu=0$\\[.2em]
        saddle-node (SN) & SS & Eq.~\eqref{eq:saddle-node-bifurcation}&-
        \\[.2em]
        drift-pitchfork (DP) & SS \& TW & Eq.~\eqref{eq:drift-pitchfork-bifurcation}& Eq.~\eqref{eq:TB_DP}
        \\[.2em]
        Hopf (H) & MW \& TW & Eq.~\eqref{eq:Hopf-bifurcation}& Eq.~\eqref{eq:TB_H}
        \\[.2em]
        drift-pitchfork  & \multirow{2}{*}{MW \& SW} & \multirow{2}{*}{numerical}& \multirow{2}{*}{Eq.~\eqref{eq:TB_DPLC}}
        \\of limit-cycle (DPLC)&&&\\[.2em]
        heteroclinic (Het) & SW \& SS & numerical& Eq.~\eqref{eq:TB_Het}\\[.2em]
        Saddle-node & \multirow{2}{*}{SW \& SS} & \multirow{2}{*}{numerical}&\multirow{2}{*}{-}\\
        infinite period (SNIPer) &&&
    \end{tabular}
    \label{tab:bifurcation-conditions}
\end{table}

In this reduced ODE system, the SS and TW correspond to two different types of fixed points, namely, with $\sin \eta = 0$ and $\sin \eta \neq 0$, respectively.
The existence and stability conditions for these states can be derived by performing a local bifurcation analysis as detailed in Appendices~\ref{app:Turing-wave} and \ref{app:local-bif-derivation}.
The results are summarized in Table~\ref{tab:bifurcation-conditions}.

In addition to the O(2)-symmetry, the considered NRSH model \eqref{eq:NRSH} and its one-mode approximation \eqref{eq:nr_AE} feature discrete internal symmetries, which relate different parameter configurations and can be used to simplify the analysis.
In particular, we may restrict ourselves to an anti-aligning reciprocal interaction $\chi \geq 0$ as the cases $\chi < 0$ can be obtained with the symmetry transformation $(\psi,\chi,\alpha) \mapsto (-\psi,-\chi,-\alpha)$, i.e., attractive interactions between $\phi$ and $\psi$ are equivalent to repulsive interactions between $\phi$ and $-\psi$.
Further, the symmetry transformation $(\phi, \psi, \alpha, \delta) \mapsto (\psi, \phi, -\alpha, -\delta)$ allows us to focus on the case $\alpha\geq 0$.
In other words, we may reorder the two fields, such that $\phi$ is always the \enquote{chased} species and $\psi$ the \enquote{chasing} one.
Finally, we set $\chi = 1$ for all numerical simulations.
Note that as long as $\chi \neq 0$, this does not restrict the generality as it corresponds to the choice of a reference timescale, i.e., all other parameters can be seen as given in units of $\chi$.
For all simulations of the NRSH model \eqref{eq:NRSH}, we set $k_{\mathrm{c}} = 1$.

\section{Phase diagram}\label{sec:phase}
\begin{figure*}
    \centering
    \includegraphics[width=\textwidth]{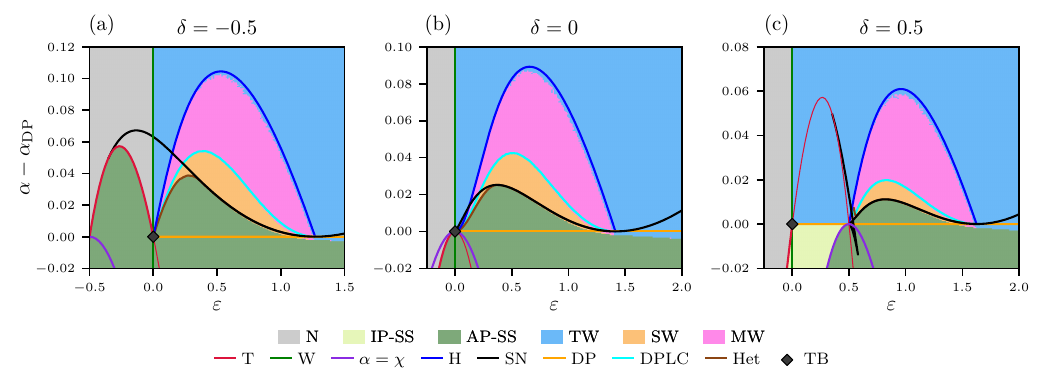}
    \caption{
        Phase diagram for (a) $\delta = -0.5$, (b) $\delta = 0$, and (c) $\delta = 0.5$ obtained by numerical calculation of the NRSH~\eqref{eq:NRSH} at $\chi = 1$.
        The numerical scheme and criteria for phase classification, including the distinction between in-phase (IP-SS) and anti-phase (AP-SS) steady states, are described in Appendix~\ref{app:methods}.
        The theoretical phase boundaries corresponding to the local or global bifurcation curves obtained with the one-mode approximation \eqref{eq:nr_AE} are also overlaid on the phase diagram.
        Black diamonds mark the location of the codimension-two Takens-Bogdanov bifurcation.
        The derivation of the theoretical curves is described in Appendices~\ref{app:Turing-wave} and \ref{app:local-bif-derivation}.
    }
    \label{fig:PDE-phase-diagram}
\end{figure*}

This section presents an overview of the stable states in the $(\varepsilon, \alpha)$-parameter plane for $\delta = - 0.5$, $0$ and $0.5$.
The resulting phase diagrams given in Fig.~\ref{fig:PDE-phase-diagram} are determined by time simulations: for each combination of $\alpha$ and $\varepsilon$, the NRSH model~\eqref{eq:NRSH} is initialized in the Null state with small noise, and the resulting phase (see Fig.~\ref{fig:sptm-PDE-trajectories-ODE}) is identified after a long transient.
Details of the numerical scheme and the criteria used to distinguish the phases, including the distinction between in-phase (IP-SS) and anti-phase (AP-SS) steady states, are given in Appendix~\ref{app:methods}.
Note that Fig.~\ref{fig:PDE-phase-diagram}(b) reproduces the result from a previous study on the symmetric case $\delta = 0$ in  Ref.~\onlinecite{TIKK2024pre}.
The phase diagrams are superimposed with semi-analytical and numerical phase boundaries obtained from the bifurcation analysis of the one-mode approximation~\eqref{eq:nr_AE} (see Table~\ref{tab:bifurcation-conditions}).
As verification, the drift-pitchfork, Hopf, and saddle-node bifurcations are also numerically computed in the full NRSH model~\eqref{eq:NRSH}.
As expected, deviations from the one-mode approximation are minimal and only visually distinguishable for large $\varepsilon$ (not shown).
Note that the nonreciprocal coupling strength $\alpha$ on the vertical axis is given relative to the nonreciprocal coupling strength $\alpha_{\mathrm{DP}}$ at the drift-pitchfork bifurcation (see Eq.~\eqref{eq:drift-pitchfork-bifurcation}).

Starting from the top left quadrant of all phase diagrams in Fig.~\ref{fig:PDE-phase-diagram}, where the Null state is stable, two distinct instabilities mark the onset of pattern formation.
On the one hand, a primary Turing (T) bifurcation gives rise to the SS phase at $\varepsilon = - \sqrt{\chi^2 + \delta^2 - \alpha^2} < 0$ (see Eq.~\eqref{eq:Turing-bifurcation}).
For $\delta < 0$, the primary Turing bifurcation can become subcritical, resulting in a small region of bistability between the Null state and the SS (see Appendix~\ref{app:Turing-wave} for details).
Although AP-SS alignment is preferred in the reciprocal case due to $\chi > 0$, both AP-SS and IP-SS alignment can occur when nonreciprocal interactions are present.
However, stable IP-SS alignment is only observed at the onset of motion in a small region of Fig.~\ref{fig:PDE-phase-diagram}(c).

On the other hand, for large nonreciprocal interaction strength $\alpha^2>\chi^2+\delta^2$, a primary wave (W) bifurcation occurs at $\varepsilon = 0$.
Here, a TW and an SW bifurcate simultaneously.
In our model, these two states always bifurcate supercritically, i.e., in the positive $\varepsilon$-direction.
Furthermore, TW are always stable, and SW are always unstable in the vicinity of the wave bifurcation, a finding confirmed analytically in Appendix~\ref{app:Turing-wave}.
Within the one-mode approximation-\eqref{eq:nr_AE} the phase difference of right-traveling waves is restricted to $\eta\in[\pi,2\pi]$, i.e., the chasing species $\psi$ is always closer to catching up with the chased species $\phi$ than vice versa.
A special case is given by $\chi \varepsilon = \alpha \delta$, where the phase shift is $\eta=3\pi/2$.

With these two primary bifurcations as a prerequisite, we turn to the transitions in the nonlinear regime, the primary focus of this study.
Since the SS phases exist at small nonreciprocal interaction strength $\alpha$, whereas the TW phases are observed for large $\alpha$, the question arises: how do SS transition into TW in the nonlinear regime, i.e., which are the possible routes to traveling waves and under which conditions do they occur?
Figure~\ref{fig:PDE-phase-diagram} reveals that these transitions involve stable MW and SW states at intermediate $\alpha$.
Specifically, either a direct route from AP-SS to TW is observed, or a route with several intermediates: AP-SS $\rightarrow$ SW $\rightarrow$ MW $\rightarrow$ TW.
For $\delta \leq 0$, the latter, more intricate route occurs directly at the onset of pattern formation and is replaced by the former simpler route at large $\varepsilon$.
In contrast, for $\delta > 0$, a direct transition from IP-SS to TW is observed for $\varepsilon \in [0, \delta]$, followed by an intermediate $\varepsilon$ range with the AP-SS $\rightarrow$ SW $\rightarrow$ MW $\rightarrow$ TW route, before finally at large $\varepsilon$ the simple AP-SS $\rightarrow$ TW route appears.

\begin{figure*}[htbp]
    \centering
    \includegraphics[width=\textwidth]{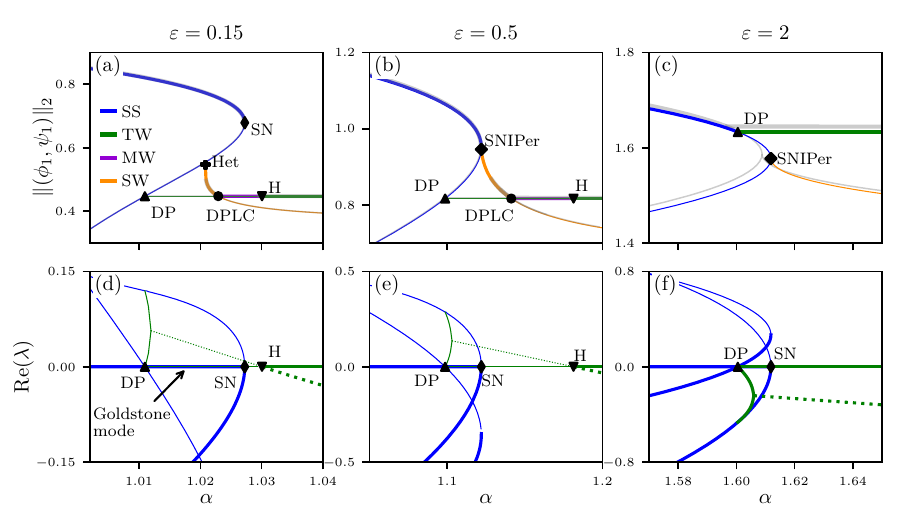}
    \caption{
    Three different bifurcation scenarios in the NRSH model~\eqref{eq:NRSH} with nonreciprocal coupling strength $\alpha$ as continuation parameter for varying values of $\varepsilon$ for the transition from steady states (SS) to traveling waves (TW) in the symmetric case $\delta = 0$.
    Panels (a) to (c) show bifurcation diagrams calculated in the one-mode approximation~\eqref{eq:nr_AE} with the root-mean-square amplitude $\|(\phi_1,\psi_1)\|_2$ in Eq.~\eqref{eq:norm_one_mode} as a solution measure.
    Stable [unstable] states are indicated by thick [thin] lines.
    Gray lines indicate the same bifurcation diagrams obtained with the full NRSH model with spatiotemporal $L^2$-norm~\eqref{eq:norm_NRSH}.
    Panels (d) to (f) show the growth rates $\mathrm{Re}(\lambda)$ of the relevant eigenvalues for the SS (blue) and TW state (green), for the latter calculated in the co-moving frame.
    Solid [dotted] lines indicate real [complex] eigenvalues.
    The SS and TW states spontaneously break the continuous O(2)-symmetry of Eq.~\eqref{eq:NRSH}, resulting in a permanently present Goldstone zero mode.
    }
    \label{fig:bif-diagrams}
\end{figure*}

To further elucidate the different routes to traveling states, one-parameter bifurcation diagrams are numerically determined using $\alpha$ as a control parameter, see Figs.~\ref{fig:bif-diagrams}(a)--(c) (Appendix~\ref{app:methods} gives the numerical details).
We focus on the symmetric case $\delta = 0$ and choose $\varepsilon = 0.15$, $0.5$ and $2$.
Again, the accuracy of the one-mode approximation is confirmed, as only slight deviations from the full NRSH model are observed (only visible at $\varepsilon = 2$ in Fig.~\ref{fig:bif-diagrams}(c), gray lines).
Additionally, Figs.~\ref{fig:bif-diagrams}(d)--(f) give the relevant eigenvalues of the SS and TW states.
In all three cases, as $\alpha$ increases, the transition starts with a drift-pitchfork bifurcation (DP).
Here, an eigenmode of the SS state coalesces with the Goldstone zero mode of translation.
Consequently, a TW state bifurcates from the SS~\footnote{This mode coalescence was also recently discussed in Ref.~\onlinecite{FHLV2021n} (there dubbed an \enquote{exceptional transition}).
However, our picture of eigenvalues differs from Fig.~1h in Ref.~\onlinecite{FHLV2021n}, as we distinguish whether the eigenvalues refer to the SS or TW phase and clearly indicate the permanently present Goldstone mode for both SS and TW phases.}.

Next, we highlight a crucial difference distinguishing the intricate SS $\rightarrow$ SW $\rightarrow$ MW $\rightarrow$ TW route from the SS $\rightarrow$ TW case: for the former (Figs.~\ref{fig:bif-diagrams}(a) and (b)), a branch of unstable TW bifurcates from the unstable, lower part of the SS branch, whereas in Fig.~\ref{fig:bif-diagrams}(c) a branch of stable TW bifurcates from the stable upper part of the SS branch.
In both cases, two real eigenvalues of the bifurcating TW states coalesce shortly after the drift-pitchfork bifurcation, becoming a pair of complex conjugate eigenvalues (note that this does not correspond to a bifurcation as the real part remains nonzero).
In Figs.~\ref{fig:bif-diagrams}(d) and (e), the real part of this complex pair decreases with increasing $\alpha$.
Consequently, it eventually passes zero, stabilizing the TW phase in a Hopf bifurcation.
This Hopf bifurcation gives rise to a stable branch of MW states that bifurcates supercritically in the negative $\alpha$-direction.
The branch of MW then terminates in a supercritical drift-pitchfork bifurcation of limit cycles on the branch of SW, thereby stabilizing it.
For the symmetric case $\delta = 0$, two other scenarios are found in which the branch of SW ends on the SS branch: at small $\varepsilon$, the SW orbit collides in phase space with the unstable SS fixed points in a heteroclinic bifurcation.
With increasing $\varepsilon$, this heteroclinic bifurcation approaches the saddle-node bifurcation of the SS branch, reaching it at $\varepsilon\approx 0.365$, where the heteroclinic bifurcation transforms into a SNIPer bifurcation.
Note that this SNIPer bifurcation persists in Fig.~\ref{fig:bif-diagrams} (c), i.e., after the drift-pitchfork bifurcation has passed the saddle-node and moved onto the stable upper SS phase at $\varepsilon = \sqrt{2}$.
Then, the branch of unstable SW ends at the SNIPer at the saddle-node bifurcation, which now connects two unstable branches.
Finally, we note that the difference between the two routes with distinct global bifurcations in Figs.~\ref{fig:bif-diagrams}(a) and (b) has direct consequences for possible parameter ranges of bistability.
While the route with the heteroclinic bifurcation permits small regions where the branch of SS may be bistable with the SW, MW, or TW branches, this cannot occur in the route with the SNIPer bifurcation.

From Figs.~\ref{fig:PDE-phase-diagram}, \ref{fig:bif-diagrams}, we can see that for any $\delta$, if $\varepsilon > 0$ and the nonreciprocity $\alpha$ is sufficiently large, the TW phase is realized stably and universally.
While numerical techniques reveal several different routes from SS to TW, they alone cannot explain why these particular routes exist and how they are interconnected.
To understand the universal origin of these routes and the complex spatiotemporal patterns involved, we must look beyond phenomenology and analyze the \enquote{organizing centers} of the dynamics.



\section{Takens-Bogdanov bifurcation with O(2)-symmetry organizes onset of pattern formation}\label{sec:TB}
The phase diagrams in Fig.~\ref{fig:PDE-phase-diagram} showcase the rich set of static and dynamic patterns described by the NRSH model~\eqref{eq:NRSH}.
However, the complete picture of the relations between the five observed phases (N, SS, SW, MW, TW) and of their transitions under change of parameters remains elusive.

In this section, we unveil the unifying structures underlying this complexity and demonstrate that a large part of the intricate bifurcation structures governing the transitions is organized by a single higher-codimension bifurcation.
Specifically, this point is identified as a Takens-Bogdanov (TB) bifurcation with O(2)-symmetry~\cite{DaKn1987ptrslsapes}, i.e., as the singularity where the onset of a static periodic pattern (Turing bifurcation) and the onset of traveling wave (wave bifurcation) transform into each other.
Applying normal form analysis, we show how this TB bifurcation organizes the entire spatiotemporal dynamics and the transitions involved in the change from steady to oscillatory phases.
This perspective facilitates a comprehensive understanding of the observed rich and complicated bifurcation and phase structure of \enquote{nonreciprocal phase transitions} within the universal framework of bifurcation theory.

\subsection{Normal form analysis at the Takens-Bogdanov bifurcation with O(2)-symmetry}
Leveraging established knowledge of bifurcation theory, normal form analysis is applied to the ODE system~\eqref{eq:nr_AE} obtained by the one-mode approximation close to the TB bifurcation occurring at $\varepsilon = 0$ and $\alpha = \sqrt{\chi^2 + \delta^2}$.
The goal is to derive the simplest possible system, i.e., the normal form, that describes the essential and universal dynamics close to the bifurcation point.
To do so, linear and weakly nonlinear variable transformations are used that preserve the system's O(2)-symmetry.

According to Dangelmayr and Knobloch~\cite{DaKn1987ptrslsapes}, the normal form for the Takens-Bogdanov (TB) bifurcation with O(2)-symmetry is given in terms of two complex amplitudes, $z$ and $w$ as
\begin{subequations}
    \begin{align}
        \dot{z} &= w,
        \label{eq:TB-normal-z}
        \\
        \dot{w} &= \mu z + \nu w + \left[
            A |z|^2 + B |w|^2
            + C (z \overline{w} + \overline{z} w)
        \right] z
        + D |z|^2 w.
        \label{eq:TB-normal-w}
    \end{align}\label{eq:TB-normal}
\end{subequations}
Here, $\mu$, $\nu \in \mathbb{R}$ are unfolding parameters, which measure the \enquote{distance} from the TB point at $(\mu, \nu) = (0, 0)$, an overline indicates a complex conjugate, and $A$, $B$, $C$, $D$ are real-valued coefficients.
They control the nonlinearities of the system, which determine the qualitative bifurcation behavior.
As demonstrated in Ref.~\onlinecite{DaKn1987ptrslsapes}, this normal form of a TB bifurcation with O(2)-symmetry allows for 29 different bifurcation scenarios depending on the sign of $A$ and the value of $D / M$, where $M \coloneqq 2 C + D$.

We perform linear and near-identity (i.e., weakly nonlinear) transformations on Eq.~\eqref{eq:nr_AE}, as detailed in Appendix~\ref{app:nit}.
As a result, we explicitly relate the unfolding parameters $\mu$, $\nu$ and the coefficients $A$, $B$, $C$, $D$ to the physical parameters $\varepsilon, \delta, \chi$ and $ \alpha$ of the original NRSH model~\eqref{eq:NRSH}.
These relationships are given by:
\begin{equation}
    \mu =  \delta^2 - \varepsilon^2 + \chi^2- \alpha^2,
    \quad
    \nu = 2 \varepsilon
    \label{eq:mu-nu}
\end{equation}
and
\begin{subequations}
    \begin{align}
        A &= \frac{3 \left(2 \alpha^{2} \varepsilon - \alpha \chi \delta + 2 \alpha \delta \varepsilon - \chi^{2} \varepsilon - \chi \delta^{2} - \chi \varepsilon^{2} + 2 \varepsilon^{3}\right)}{\left(\alpha + \delta\right)^{2}},
        \label{eq:A-NRSH}
        \\
        B &= - \frac{6 \varepsilon}{\left(\alpha + \delta\right)^{2}},
        \\
        C &= D = - \frac{3 \left(\alpha^{2} + \alpha \delta - \chi \varepsilon + \varepsilon^{2}\right)}{\left(\alpha + \delta\right)^{2}},
        \\
        M &= 2 C + D = 3 D.
    \end{align}\label{eq:A~M-NRSH}
\end{subequations}

A notable feature of the derived normal form coefficients \eqref{eq:A~M-NRSH} for the one-mode approximation of the NRSH model is that $C = D$ holds universally, regardless of the specific values of the physical parameters $\varepsilon$, $\delta$, $\chi$, and $\alpha$.
This further implies a fixed ratio $D / M = 1 / 3$.
According to the classification by Dangelmayr and Knobloch~\cite{DaKn1987ptrslsapes}, this algebraic constraint strongly restricts the possible bifurcation scenarios from 29 theoretically possible ones to just two: Type $\text{II}_-$ (when $A < 0$) and Type $\text{III}_-$ (when $A > 0$) in the classification of Ref.~\onlinecite{DaKn1987ptrslsapes}.
Thus, the bifurcation scenario is determined solely by the sign of the coefficient $A$.

Evaluating $A$ at the TB point yields $A = - 3 \chi \delta / (\delta + \sqrt{\chi^{2} + \delta^{2}})$.
The sign of $A$ is thus solely controlled by our physical parameter $\delta$, which represents the asymmetry of the local potentials of the two fields, and here acts as a toggle for the entire topology of the bifurcation structure when unfolded around the organizing center.
Further, to linear order, the SS solutions $(w, z) = (0, z_0)$ in the normal form~\eqref{eq:TB-normal} transform into $(\phi_0, \psi_0) \sim (1, \delta/ (\chi + \sqrt{\chi^2 + \delta^2})) z_0$ in the one-mode approximation \eqref{eq:nr_AE}, i.e., only AP-SS [IP-SS] alignment appears in the vicinity of the TB point for $\delta < 0$ [$\delta > 0$].

\subsection{Numerical comparison}
\begin{figure}
    \centering\includegraphics[width=\linewidth]{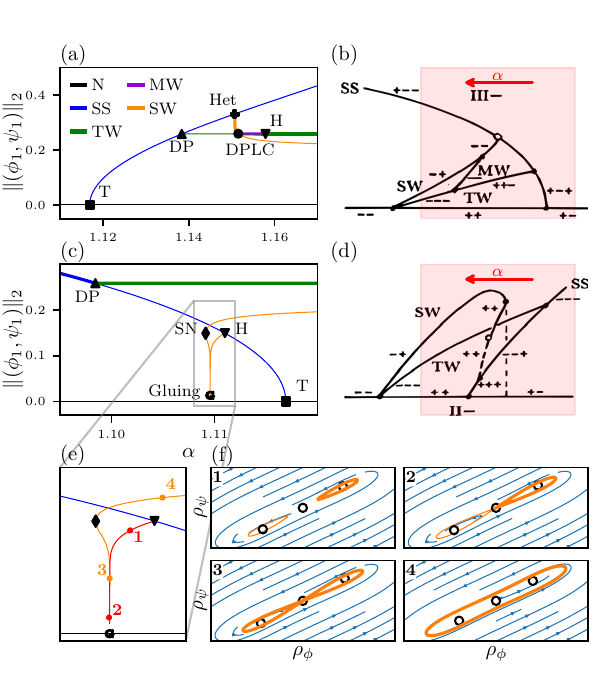}
    \caption{The two realized bifurcation scenarios in the vicinity of the O(2)-Takens-Bogdanov bifurcation.
    Panels (a) and (c) show numerical bifurcation diagrams for $\varepsilon = 0.05$ at $\delta = -0.5$, and $\delta = 0.5$, respectively, as obtained with the one-mode approximation~\eqref{eq:nr_AE}.
    Stable [unstable] states are indicated as thick [thin] lines.
    Panels (b) and (d) show the corresponding conceptual bifurcation diagrams $\text{III}_-$ ($A > 0$) and  $\text{II}_-$ ($A < 0$) occurring in the universal unfolding in Ref.~\onlinecite{DaKn1987ptrslsapes}, with plus and minus signs indicating the number of stable and unstable eigenvalues of the respective branches (used with permission of the Royal Society (U.K.), from Ref.~\onlinecite{DaKn1987ptrslsapes}; permission conveyed through Copyright Clearance Center, Inc.).
    The red shading indicates the bifurcations that are accessible with the nonreciprocal coupling strength $\alpha$ as a bifurcation parameter (the direction of panels (b) and (d) corresponds to $\alpha$ increasing from right to left, as indicated by the red arrow).
    Panel (e) shows a magnification of the successive saddle-node, gluing, and Hopf bifurcations that terminate the standing wave branch.
    Subpanels 1 and 2 [3 and 4] of (f) show the asymmetric [symmetric] SW limit cycles in the amplitude representation of the phase space~\eqref{eq:phase_difference}.
    The hollow circles represent the unstable fixed points corresponding to N and SS states.
    }
    \label{fig:bt_scenarios}
\end{figure}

Numerical simulations corroborate these two scenarios and elucidate the emerging bifurcation structure, as shown in Fig.~\ref{fig:bt_scenarios}.
Specifically, Figs.~\ref{fig:bt_scenarios}(a) and \ref{fig:bt_scenarios}(c) represent bifurcation diagrams obtained with the one-mode approximation \eqref{eq:nr_AE} using $\alpha$ as control parameter at fixed $\varepsilon = 0.05$, i.e., vertical cuts through Figs.~\ref{fig:PDE-phase-diagram}(a) and \ref{fig:PDE-phase-diagram}(c) in the immediate vicinity of the TB point.
Figs.~\ref{fig:bt_scenarios}(b) and \ref{fig:bt_scenarios}(d) show the corresponding conceptual bifurcation structures derived in Ref.~\onlinecite{DaKn1987ptrslsapes}, where the plus [minus] signs indicate the number of unstable [stable] eigenvalues of the respective branches.
Note that an increase in their bifurcation parameters corresponds to a clockwise rotation around the TB point in Fig.~\ref{fig:PDE-phase-diagram} starting from the stable Null state that appears in the upper left quadrant.
Consequently, the conceptual diagrams include the wave bifurcation, a feature absent in our present numerical analysis.

The Type~$\text{III}_-$ scenario for $\delta < 0$ (upper row of Fig.~\ref{fig:bt_scenarios}) contains a supercritical Turing bifurcation, where an unstable SS bifurcates from the already unstable Null state, otherwise resembling the symmetric case $\delta = 0$ discussed in Section~\ref{sec:phase} [Fig.~\ref{fig:bif-diagrams}(c)].
In particular, the system exhibits the same succession of bifurcations and stability of time-dependent branches, i.e., SW $\rightarrow$ MW $\rightarrow$ TW.
Figure~\ref{fig:bt_scenarios}(a) also features a stable AP-SS of large amplitude (not shown) present for all $\alpha$-values.
However, capturing such a large-amplitude state is beyond the scope of the leading-order normal form.

The Type~$\text{II}_-$ scenario for $\delta > 0$ (second row of Fig.~\ref{fig:bt_scenarios}) exhibits less intricacy in terms of stable states.
Here, an unstable SS bifurcates subcritically from the unstable Null state, undergoes a Hopf bifurcation, and finally stabilizes in a drift-pitchfork bifurcation.
The latter is supercritical and gives rise to a stable TW state.
The normal form analysis also clarifies why the MW and SW states are not observed in time simulations for $\delta > 0$ in the vicinity of the TB point.
While the MW state does not exist, the SW states exist but remain unstable.
As shown in the magnification in Fig.~\ref{fig:bt_scenarios}(e), they originate from the Hopf bifurcation on the SS branch, which gives rise to two asymmetric SW states related by inversion symmetry, each circling one of the two IP-SS fixed points.
These limit cycles then collide with the fixed point at the origin (Null state) in a gluing bifurcation\cite{GaGT1988n} and form a symmetric SW state that circles all three mentioned fixed points.
Note that the spatiotemporal $L^2$-norm approaches zero at the gluing bifurcation, as the SW states spend the majority of their diverging period at $(0,0)$.
Subsequently, the symmetric SW branch undergoes a saddle-node bifurcation of limit cycles, where it folds back toward larger $\alpha$.

\begin{figure}
    \centering
    \includegraphics[width=\linewidth]{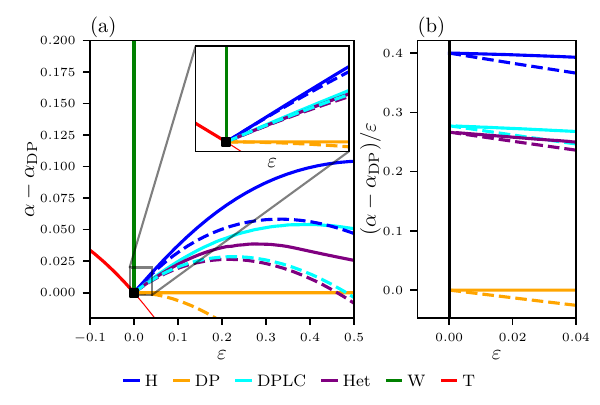}
    \caption{
    Quantitative evaluation of the predictions of the normal form of the O(2)-symmetric Takens-Bogdanov bifurcation. Panel (a) for $\delta = -0.5$ compares the loci of bifurcations in the $(\varepsilon, \alpha)$-plane computed with the normal form~\eqref{eq:TB-normal} (dashed lines) and with the one-mode approximation \eqref{eq:nr_AE} (solid lines).
    Panel (b) shows the slope $(\alpha - \alpha_{\mathrm{DP}}) / \varepsilon$ in the vicinity of the TB point, proving that the corresponding lines of codimension-one bifurcations determined on the two levels of description enter the TB point tangentially to each other.
    }
   \label{fig:tb_quantitative}
\end{figure}

Finally, a quantitative comparison relates the predictions obtained with the normal form \eqref{eq:TB-normal} and with the one-mode approximation \eqref{eq:nr_AE}.
Figure~\ref{fig:tb_quantitative} presents the computed two-parameter diagrams for $\delta = - 0.5$ on the two levels of description.
The formulas giving the theoretical curves for all codimension-one bifurcation lines that emerge from the codimension-two TB point are given in Eqs.~\eqref{eq:TB_H}--\eqref{eq:TB_DPLC} in Appendix~\ref{app:global-bif-results}.
As expected, the normal form theory provides accurate quantitative predictions in the immediate vicinity of the TB point; all bifurcation lines in Eqs.~\eqref{eq:nr_AE} (solid lines) enter the TB point tangentially to the corresponding line from the normal form (dashed lines).
This is further illustrated in Fig.~\ref{fig:tb_quantitative}(b), which shows the numerical slopes $(\alpha - \alpha_{\mathrm{DP}}) / \varepsilon$ calculated close to the TB point.
Note that the deviations become significant already for relatively small values of $\varepsilon$.
This is especially pronounced for the drift-pitchfork bifurcation (orange lines).

\section{Codimension-two bifurcations organize routes to traveling states}\label{sec:param_space}

\begin{figure*}[htb]
    \centering
\includegraphics[width=\linewidth]{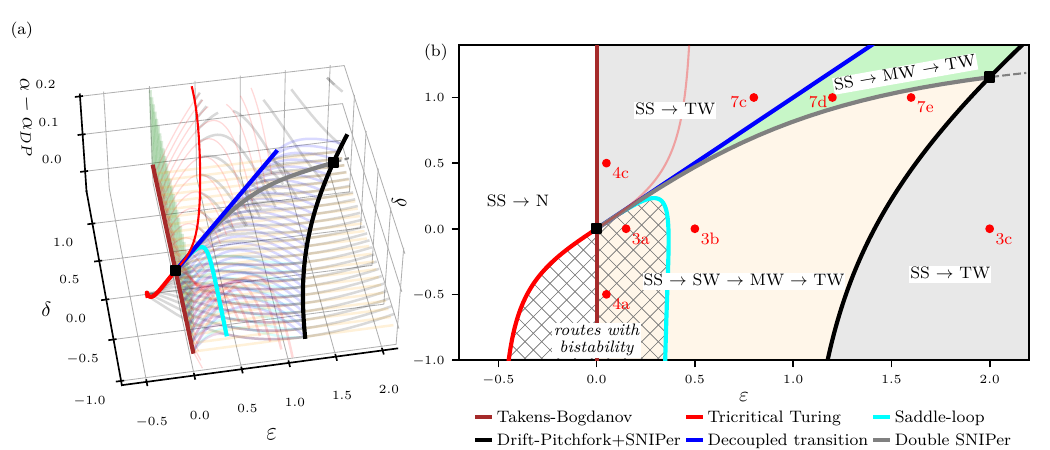}
    \caption{
    Organization of the parameter space of the NRSH model~\eqref{eq:NRSH}: Panel (a) shows the lines of the codimension-two bifurcations listed in Table~\ref{tab:codim2_bifurcation-conditions} as heavy lines in the parameter space spanned by $\varepsilon$, $\delta$, and $\alpha$.
    Thin colored lines indicate the surfaces formed by the loci of the codimension-one bifurcations listed in Table~\ref{tab:bifurcation-conditions} (colors as in Fig.~\ref{fig:PDE-phase-diagram}).
    The thick lines in panel~(b) give the projection of the lines of codimension-two bifurcations onto the $(\varepsilon, \delta)$-plane.
    Panel~(b) also indicates the resulting qualitatively different routes from SS to TW ($\varepsilon > 0$) and from SS to N ($\varepsilon < 0$).
    The filled red circles indicate the parameter values corresponding to the one-parameter bifurcation diagrams in Figs.~\ref{fig:bif-diagrams}, \ref{fig:bt_scenarios}, and \ref{fig:delta_plus_scenarios}, as indicated.
    }
    \label{fig:param_space}
\end{figure*}

Now, with the background of the semi-analytical and numerical observations in Section~\ref{sec:phase} and the rigorous mathematical analysis in the vicinity of the TB point in Section~\ref{sec:TB}, we return to the overall question of possible routes to traveling states.
That is, in contrast to Section~\ref{sec:TB}, here, we also consider strongly nonlinear regimes accessible with the one-mode approximation~\eqref{eq:nr_AE}.
In particular, for every choice of $\delta$ and $\varepsilon$, we determine the sequence of stable states that is traversed as the nonreciprocal coupling strength $\alpha$ increases. Qualitatively, one can distinguish different routes by considering the projections of a few codimension-two bifurcation manifolds onto the $(\delta, \varepsilon)$-plane, which is illustrated in Fig.~\ref{fig:param_space}.
With the exception of the saddle-loop bifurcation, they obey the simple analytic expressions listed in Table~\ref{tab:codim2_bifurcation-conditions} with derivation details given in Appendix~\ref{app:local-bif-derivation}.

\begin{table}[htbp]
    \centering
    \caption{
    Dynamically relevant codimension-two bifurcations.
    With the exception of the tricritical Turing bifurcation, the necessary conditions are parametrized by $\chi$ and $\delta$, i.e., solved for $\varepsilon$ and $\alpha$.
    }
    \begin{tabular}{c||c}
        Bifurcation &  Conditions
        \\
        \hline
        \multirow{2}{*}{Takens-Bogdanov} & $\varepsilon=0$
        \\
        &$\alpha=\sqrt{\chi^2+\delta^2}$
        \\ \hline
        \multirow{2}{*}{Tricritical Turing}&$(\chi+\alpha)^2(\varepsilon-\delta)+(\varepsilon+\delta)^3=0$
        \\
        & $\varepsilon^2-\delta^2-\chi^2+\alpha^2=0$
        \\ \hline
        Saddle-loop & \multirow{2}{*}{numerical (shooting)}
        \\
        (SNIPer + heteroclinic)&\\\hline

        Drift-Pitchfork&$\varepsilon=\frac{1}{\sqrt{2}}\sqrt{4\chi^2+\delta^2+\delta\sqrt{4\chi^2+\delta^2}}$
        \\
        +SNIPer (+Hopf)&$\alpha=\frac{1}{\sqrt{2}}\sqrt{4\chi^2+\delta^2-\delta\sqrt{4\chi^2+\delta^2}}$
        \\ \hline
        \multirow{2}{*}{Decoupled transition} & $\varepsilon = \delta$
        \\
        &$\alpha=\chi$\\\hline

        Double Saddle-node &$\varepsilon = \frac{\sqrt{2}\delta\chi}{\sqrt{2\chi^2-\delta^2}}$
        \\
        (Double SNIPer) &$\alpha =\frac{4\chi^2-\delta^2}{2\sqrt{2}\sqrt{2\chi^2-\delta^2}}$
        \\ \hline

    \end{tabular}
    \label{tab:codim2_bifurcation-conditions}
\end{table}

Besides the O(2)-symmetric TB bifurcation discussed in Section~\ref{sec:TB}, we identify five further codimension-two bifurcations that involve stable states.
First, for $\varepsilon < 0$, there exists a tricritical Turing bifurcation (red line), where the emerging branch of SS changes from super- to sub-critical. It encloses a $(\delta,\varepsilon)$-region, where the transition from SS to the trivial solution N exhibits hysteresis [cf.\ Fig.~\ref{fig:PDE-phase-diagram}(a)].
In this parameter region, bistability may result in snaking branches of steady, spatially localized states on larger domains in the NRSH model~\eqref{eq:NRSH}.
Note that this effect is induced by dominant nonreciprocal coupling $\alpha>\chi$, since the individual decoupled SH equations with the used simple cubic nonlinearity always show supercritical behavior.
Further, subcritical behavior can only appear for $\delta < \varepsilon < 0$ (cf.\ Appendix~\ref{app:Turing-wave}), i.e., only if the uniform state of the decoupled chased species is stable ($\varepsilon + \delta < 0$) and the decoupled chasing species is pattern-forming ($\varepsilon - \delta > 0$).

Second, there exists a saddle-loop bifurcation (light-blue line), where the heteroclinic bifurcation changes to a SNIPer bifurcation.
It plays a similar role as the tricritical Turing with respect to the occurrence of hysteresis but for the oscillatory phases.
The largest part of this bifurcation manifold is almost independent of $\delta$; thus, hysteretic transitions with bistability between SS and SW, MW or TW states appear for small values of $\varepsilon \lessapprox 0.365$.
However, in contrast to the bistability of N and SS states for $\varepsilon<0$, bistability also appears in a very small parameter region for positive $\delta \lessapprox  0.24$.

Third, a bifurcation formed by coinciding drift-pitchfork and SNIPer bifurcation (black line) terminates the existence region of stable SW and MW phases in positive $\varepsilon$ direction.
As discussed in Section~\ref{sec:phase}, the drift-pitchfork bifurcation passes through the saddle-node and moves from the lower, unstable SS branch to the upper stable SS branch, such that afterwards the TW branch bifurcates already stable for $\varepsilon > \varepsilon_{\mathrm{DP+SNIPer}}$.
As can be understood from Figs.~\ref{fig:bif-diagrams}(e) and \ref{fig:bif-diagrams}(f), the Hopf bifurcation on the TW branch also ceases to exist at this codimension-two bifurcation, as the complex conjugate modes of the emerging TW branch change from unstable to stable.
We also emphasize that, even though this bifurcation is called a \enquote{tetracritical point} in Ref.~\onlinecite{FHLV2021n} due to the four different phases in its vicinity, it is a generic codimension-two bifurcation.
In particular, it is structurally stable, i.e., it does not qualitatively change if nonlinear couplings are introduced (see the discussion in Ref.~\onlinecite{KnMo1990pra}, where a normal form is proposed that contains all participating local bifurcations).

\begin{figure}
    \centering
    \includegraphics[width=\linewidth]{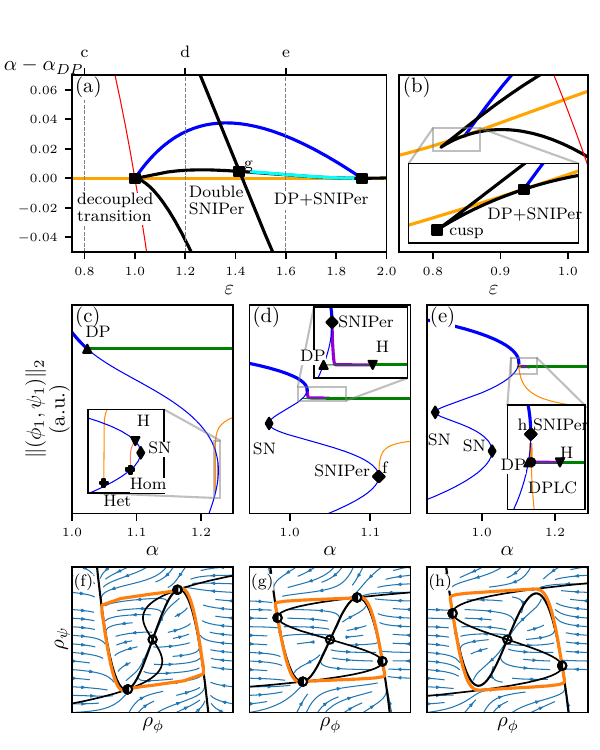}
    \caption{Transition scenarios for $\delta=1$: Panel (a) shows a two-parameter bifurcation diagram centered at the three organizing codimension-two bifurcations in the nonlinear regime (indicated by squares).
    Panel (b) shows the structural instability of the decoupled transition, i.e., its unfolding into two generic codimension-two points with the addition of the nonlinear interaction terms $\beta \phi \psi^2$ and $\beta \phi^2 \psi$ in the NRSH model~\eqref{eq:NRSH}, here for $\beta=0.1$.
    Panels (c)--(e) show the transition scenarios to traveling waves in the different regimes separated by the codimension-two points, i.e., one-parameter bifurcation diagrams in $\alpha$.
    Panels (f)--(h) show the transition of the local phase space in the amplitude-representation~\eqref{eq:phase_difference} at the double SNIPer bifurcation, where the termination of SW changes from in-phase to anti-phase SS.
    Semi-filled circles indicate saddle-nodes that emerge where the two-cubic nullclines (black lines) are tangential.
    Orange lines indicate the terminating limit cycle.
    }
    \label{fig:delta_plus_scenarios}
\end{figure}

Fourth, we find the \enquote{decoupled transition} at $\varepsilon=\delta>0$ and $\alpha=\chi$ (dark blue line).
We name this codimension-two bifurcation after the latter condition, which is decisive for the bifurcation structure, i.e., that the chasing species $\psi$ decouples from the chased species $\phi$ as the attractive reciprocal interaction strength $\chi$ is exactly compensated by the nonreciprocal interaction strength $\alpha$.
In consequence, at $\alpha=\chi$, Eq.~\eqref{eq:psi_1} reduces to
\begin{equation}
    \dot{\psi}_1 = (\varepsilon - \delta) \psi_1 - 3 |\psi_1|^2 \psi_1.
\end{equation}
Hence, the SS amplitude $\psi_1$ is zero for $\varepsilon<\delta$ at $\alpha=\chi$ and passing from dominant reciprocal coupling $\alpha<\chi$ to dominant non-reciprocal coupling $\alpha>\chi$ does not correspond to a bifurcation, but to a zero-crossing of $\psi_1$ resulting in a change from anti-alignment to alignment.
However, if we increase $\varepsilon$ past $\varepsilon = \delta > 0$, the chasing species $\psi$ starts to form patterns even in the decoupled case, and we find the codimension-two \enquote{decoupled transition}, which is shown on the left in the two-parameter diagram in Fig.~\ref{fig:delta_plus_scenarios}(a).
To the right of the decoupled transition, small-amplitude states of $\psi$ can either be in-phase or anti-phase with the $\phi$ pattern and also emerge from the unstable N phase in a secondary Turing bifurcation, leading to the presence of three different SS phases in the vicinity of the codimension-two bifurcation.
In consequence, for $\varepsilon>\delta$, i.e., to the right of the decoupled transition in Fig.~\ref{fig:delta_plus_scenarios}(d), two further saddle-node bifurcations appear. The drift-pitchfork bifurcation attaches to the lower, unstable part of the AP-SS, which also gives rise to the secondary Hopf bifurcation that stabilizes the TW phase.
A detailed analysis of this transition in the context of a nonreciprocal Cahn-Hilliard model is shown in Fig.~11 of Ref.~\onlinecite{FrWT2021pre}, where the $\delta>0$ scenario is called \enquote{subminus} and one-parameter bifurcation diagrams are given in terms of the self-interaction (here called $\varepsilon$).

Fifth, there exists a double SNIPer bifurcation for $\delta>0$ (gray line in Fig.~\ref{fig:param_space}) characterized by two simultaneous SNIPer bifurcations of the SW branch, see the characteristic phase-space in the amplitude representation in Fig.~\ref{fig:delta_plus_scenarios}(g).
As shown in Fig.~\ref{fig:delta_plus_scenarios}(d) and \ref{fig:delta_plus_scenarios}(e), the drift-pitchfork bifurcation always remains on the upper segment of the SS branch corresponding to AP-SS, such that the lower in-phase part of the SS branch is always unstable with respect to drift.
However, the IP-SS plays a role in the termination of the SW phase, as to the left of the double SNIPer bifurcation in Fig.~\ref{fig:delta_plus_scenarios}(a), SW states terminate at the lower, in-phase part of the SS branch as shown in panel~(f)  of Fig.~\ref{fig:delta_plus_scenarios}.
Stabilization of the SW branch in terms of the drift-pitchfork bifurcation of limit cycles is only possible to the right of the double SNIPer bifurcation, when the SW branch again terminates at the AP-SS branch (Fig.~\ref{fig:delta_plus_scenarios}(h)).
Thus, the decoupled transition and double SNIPer bifurcation shape the boundaries of a $(\varepsilon, \delta)$-region where the SW cannot stabilize, resulting in an SS $\rightarrow$ MW $\rightarrow$ TW route to traveling states.

The analysis of codimension-two bifurcations also allows us to reflect on the structural stability of the NRSH model~\eqref{eq:NRSH}, in particular, on its embedding into model extensions that include further nonlinear terms.
With the exception of the decoupled transition, all bifurcations are generic bifurcations in systems with O(2)-symmetry, i.e., their topology in Fig.~\ref{fig:param_space} remains unchanged if small nonlinear coupling terms are added.
In contrast, within an extended parameter space, the decoupled transition is of higher codimension, which can be intuitively understood: To decouple one species from the other, the coefficients of all coupling terms in one of the equations have to be zero, i.e., decoupling appears only in high-codimension manifolds within extended parameter spaces.
However, here we \textit{a priori} restricted ourselves to linear interactions, and in consequence, decoupling appears in the codimension-one manifold $\chi = \alpha$.
We briefly demonstrate this degeneracy in Fig.~\ref{fig:delta_plus_scenarios}(b) by adding the variational nonlinear coupling terms $\beta \phi \psi^2$ and $\beta \phi^2 \psi$ to Eqs.~\eqref{eq:NRSH-phi} and \eqref{eq:NRSH-psi}, respectively.
In the two-parameter bifurcation diagram for $\beta = 0.1$, the decoupled transition dissolves into two generic codimension-two bifurcations, here a cusp and another drift-pitchfork+SNIPer bifurcation, which also no longer coincide with the secondary Turing bifurcation on the trivial state.

Finally, we emphasize the role of the codimension-three point $(\varepsilon = 0, \delta = 0, \alpha = \chi)$ as the central organizing element for systems with linear interactions and O(2)-symmetry.
All different routes to traveling states are present in its immediate vicinity and with the exception of the DP+SNIPer bifurcation, all codimension-two bifurcations emerge from this point.
As it inherits the structural instability from the decoupled transition and is also a point of nonlinear degeneracy for the TB bifurcation, we deem an extended discussion of its unfolding an interesting avenue for future research.

\section{Discussion and conclusion}\label{sec:conc}

In this work, we have investigated nonreciprocal phase transitions in the NRSH model, focusing on the emergence of spatiotemporal patterns induced by nonreciprocal interactions.
While other studies emphasize the role of the drift-pitchfork bifurcation (as a codimension-one \enquote{exceptional transition}~\cite{FHLV2021n}) as the central organizing element, the present analysis has demonstrated that the transition from steady to oscillatory phases involves a more complex interplay of several local and global bifurcations.

The one-mode approximation, which we have used to reduce the original PDE system to two coupled complex ODEs while retaining the underlying O(2)-symmetry, shows excellent agreement with the original system.
This reduction facilitates the determination of SS and TW as fixed points as well as analytic criteria for their stability and bifurcations.
At the onset of pattern formation, the system is further reduced to the normal form of the O(2)-symmetric Takens-Bogdanov bifurcation.
This analytical step allows for a comprehensive reconstruction of the bifurcation landscape, including global bifurcations of SW and MW phases that are otherwise analytically inaccessible.
However, the comparison with numerical results indicates that this validity range is relatively limited, as deviations become quantitatively prominent already at small distances from the codimension-two point.

Notably, for nonzero potential depth differences, the normal form coefficients are non-degenerate but subject to the intrinsic algebraic constraint $D/M = 1/3$, imposed by the structure of local cubic nonlinear terms.
Consequently, the NRSH model switches between only two out of 29 possible bifurcation scenarios present in the universal unfolding depending on the sign of the potential depth difference.

Thus, stable SW and MW phases appear at the onset of pattern formation, if and only if the chased species possesses stabilizing self-interaction, like an inhibitor in a reaction-diffusion system.
Higher degeneracy at the onset of pattern formation occurs at the limiting case of vanishing potential depth difference, i.e., equal self interactions. There, the Takens-Bogdanov bifurcation additionally coincides with several other codimension-two bifurcations including the already degenerate decoupled transition.

The investigation of further codimension-two bifurcations in the one-mode approximation has extended the analysis into the fully nonlinear regime.
In particular, SNIPer bifurcations of SW, which are absent in the TB normal form, should play a crucial role in the distinct routes from SS to TW in strongly nonlinear regimes.
Compared to previous studies~\cite{TIKK2024pre}, the present analysis offers a more complete exploration of the three-dimensional parameter space of a linearly coupled O(2)-symmetric system with self-stabilizing cubic nonlinearities, focusing on physically relevant quantities such as the nonreciprocal interaction strength.

The theoretical framework presented here is applicable to various experimental systems characterized by a dominant characteristic wavelength and activity arising from nonreciprocal interactions between two species.
Examples include dynamic patterns in Min proteins~\cite{RWSW2025np}, active Rosensweig patterns~\cite{RHST2025arxiv}, atomic nanowires~\cite{asaba2023growth}, optical parametric oscillators~\cite{longhi1996swift}, or visual cortical maps~\cite{reichl2012coordinated}.
Furthermore, O(2)-symmetry in physical systems may also stem from rotational invariance in two dimensions, which links the analyzed amplitude equations to chiral active particles with nonreciprocal alignment interactions~\cite{KrKl2024prl, KrKl2022njp, FHLV2021n,KrKl2025cp}.
Although quantitative application requires external control of the nonreciprocal interaction strength and knowledge of reciprocal and self-interaction strengths---parameters that may be experimentally inaccessible---the results are well-suited for qualitative reasoning.
In particular, the observation of qualitatively distinct routes to traveling wave states provides a means to map experimental observations to specific parameter regions of the model.

This work lays the foundation for a deeper understanding of nonreciprocal phase transitions and points to various avenues for future research.
First, extending the model to a stochastic field theory would allow for the analysis of critical fluctuations and entropy-production rates for the various transitions to oscillatory phases, similar to the analysis for drift-pitchfork and wave bifurcations in Refs.~\onlinecite{SuKL2023pre,SuKL2023prl}.
Second, the NRSH model serves as a minimal framework for studying the kinetics of nonreciprocal phase transitions in spatially extended systems.
Allowing amplitudes to vary on large length scales extends the one-mode approximation to a system of nonreciprocally coupled Ginzburg-Landau-type equations, enabling the discussion of spatiotemporal structures such as fronts, defects, and localized states.
A similar analysis of the NRSH model with symmetric self-interactions, purely nonreciprocal couplings and distinct critical wavenumbers is given in Ref.~\onlinecite{BFNR2018c}, whereas related spatiotemporal chaos and defect dynamics in the complex Swift-Hohenberg equation are discussed in Refs.~\onlinecite{aranson1995domain, sakaguchi1998pattern, gelens2009faceting, gelens2011traveling}.
Here, the discovered region of bistability between steady and oscillatory phases is of particular interest, as it may give rise to several types of \enquote{nonreciprocal first-order phase transitions.}

Another aspect is the topological rigidity observed in our analysis, i.e., the fixed ratio $D / M = 1 / 3$ found in the normal form stemming from the locality of nonlinear interactions in the underlying field equations.
To access the other bifurcation topologies predicted by normal form theory, one must break this constraint, for instance, by introducing nonlocal interactions such as coupling terms involving spatial derivatives.
A comprehensive mathematical classification of such generalized bifurcation structures in the context of nonreciprocal interactions remains an intriguing direction for future research.

Finally, we comment on the connection of the present results to systems with large-scale instabilities and mass conservation described, e.g., by nonreciprocal Cahn-Hilliard models\cite{SaAG2020prx,YoBM2020pnasusa,FrTh2023prl}. There, the underlying conservation laws prevent uniform perturbations corresponding to zero-wavenumber modes, such that the critical wavenumber at the onset of pattern formation often corresponds to a wavelength of system size (type II in the classification of Cross and Hohenberg\cite{CrHo1993rmp}).
Remarkably, on the level of a one-mode approximation, nonreciprocal Cahn-Hilliard models with zero mean densities are governed by identical equations with shifted coefficients to the NRSH model here treated without external chemical potentials (see, e.g., Ref.~\onlinecite{YoBM2020pnasusa} for the one-mode approximation of a nonreciprocal Cahn-Hilliard model). This is also the case for Swift-Hohenberg models with external chemical potentials and Cahn-Hilliard models with nonzero mean densities. Therefore, one-mode approximations establish a connection between certain models of mass-conserving and non-mass-conserving dynamics close to the onset of pattern formation through the universal framework of equivariant bifurcation theory. Further away from onset, secondary modes will enter the description whose selection heavily depends on the system specifics and the presence of mass conservation. This paves the way for future comparative studies on the implications of conservation laws.

\section*{Acknowledgment}
We would like to thank the JSPS Core-to-Core Program \textit{Advanced core-to-core network for the physics of self-organizing active matter}(JPJSCCA20230002), especially for the hospitality during the workshop \textit{Self-Organizing and Evolving Active Matter} (SOEAM25) in Dresden and the associated visit by H.K.\ and Y.T.\ in Münster.
Y.T.\ acknowledges the support from the Japan Science and Technology Agency (JST) SPRING (No.~JPMJSP2109) and the Cooperative Research Program of ``Network Joint Research Center for Materials and Devices'' (No.~20255007).
H.I.\ acknowledges the support by JSPS KAKENHI Grant Number JP24K06972.
S.K.\ acknowledges the support by the National Natural Science Foundation of China (No.~12274098) and the Scientific Research Starting Foundation of Wenzhou Institute, UCAS (Grant No.~WIUCASQD2021041).
H.K.\ acknowledges the support by JSPS KAKENHI Grant Numbers JP24K22311, JP24K06978, JP25K00918, by Cooperative Research Program of ``Network Joint Research Center for Materials and Devices'' (No.~20254003), and by MEXT Promotion of Distinctive Joint Usage/Research Center Support Program Grant Number JPMXP0724020292.
We thank Kristian Blom and Florian Voss for fruitful discussions.

\appendix
\section{Numerical methods}\label{app:methods}
\subsection{Direct simulation and phase classification}\label{app:direct_sim}
Numerical integration of the NRSH model~\eqref{eq:NRSH} is performed using the method of lines.
Spatial derivatives are approximated using a second-order central difference scheme, discretizing a periodic domain of size $L = 2 \pi$ into a grid of $N = 128$ points, corresponding to a grid spacing of $\Delta x = 2\pi/128 \simeq 0.05$.
Time integration is conducted using an open-source differential
equation solver employing the ESDIRK (Explicit Singly Diagonal Implicit Runge-Kutta) method~\cite{rackauckas2017differentialequations}.
Initial conditions are generated by superimposing Gaussian white noise with a variance of $10^{-4}$ onto the Null (N) phase $(\phi,\psi) = (0, 0)$.
The main numerical procedure follows the methodology detailed in our previous work~\cite{TIKK2024pre}; however, we have extended the phase classification criteria to distinguish between in-phase (IP-SS) and anti-phase (AP-SS) steady-state configurations.

Spatiotemporal phases (N, IP-SS, AP-SS, TW, SW, MW) are classified based on the power spectrum of the spatiotemporal Fourier transform $\tilde{\phi}(k_{\mathrm{c}}, \omega)$ of the dominant spatial mode.
Phases are identified according to the following hierarchical criteria:
(i) The Null (N) phase corresponds to states where the maximal power is below a small threshold ($\max_\omega |\tilde{\phi}(k_{\mathrm{c}}, \omega)|^2 < 10^{-4}$).
(ii) Steady-States (SS) manifest as a single dominant peak at $\omega = 0$.
To distinguish between IP-SS and AP-SS configurations, the spatial Pearson correlation coefficient $C_{\phi\psi}$ between the fields $\phi(x)$ and $\psi(x)$ at the final time step is calculated.
The state is identified as IP-SS if $C_{\phi\psi} \approx 1$, and as AP-SS if $C_{\phi\psi} \approx -1$.
Note that intermediate correlation values are not observed except in the decoupled limit.
(iii) The Traveling-Wave (TW) phase is characterized by a single dominant peak at $\omega \neq 0$, where the intensity ratio of the second-largest peak to the largest peak is less than $10^{-2}$.
(iv) The Standing-Wave (SW) phase is determined by the presence of two dominant peaks with approximately equal amplitudes at frequencies $\omega_1$ and $\omega_2$, satisfying $\omega_1 + \omega_2 \approx 0$.
(v) The Modulated-Wave (MW) phase corresponds to a dynamic state with multiple frequency peaks that does not satisfy the symmetry condition of the SW phase or the single-peak condition of the TW phase.

\subsection{Numerical continuation}\label{app:continuation}
To quantify the solution amplitude in bifurcation diagrams, we define the spatiotemporal $L^2$-norm for the NRSH model~\eqref{eq:NRSH} as
\begin{equation}\label{eq:norm_NRSH}
    \|(\phi, \psi)\|_{2} = \left[\frac{1}{L T}\int_0^T \int_0^L  \left(|\phi(x, t)|^2 + |\psi(x, t)|^2\right)\mathrm{d}x \, \mathrm{d}t \right]^{1 / 2}.
\end{equation}
For the one-mode approximation~\eqref{eq:nr_AE}, we use the root-mean-square amplitude
\begin{equation}\label{eq:norm_one_mode}
    {\|(\phi_1, \psi_1)\|_{2} = \left[\frac{2}{T}\int_0^T \left(|\phi_1(t)|^2 + |\psi_1(t)|^2\right)\mathrm{d}t \right]^{1 / 2}}.
\end{equation}
Note that a factor of two is introduced in \eqref{eq:norm_one_mode} to match the norm for spatially harmonic profiles on the two levels of description.

To numerically obtain the bifurcation diagrams, we employ the path continuation toolbox \textit{pde2path} \cite{UeWR2014nmma,Ueck2019ccp}. Traveling waves are obtained as steady states of the NRSH model~\eqref{eq:NRSH} in the comoving coordinates $\tilde{x}=x-ct$, i.e., we replace $\partial_t \rightarrow \partial_t -c \partial_{\tilde{x}}$ in Eqs.~\eqref{eq:NRSH}. The traveling wave velocity $c$ then acts as a Lagrange multiplier for the motion along the group orbit of the continuous translation symmetry, which we restrict via a suitable phase condition. Similarly, in the one-mode approximation \eqref{eq:nr_AE}, traveling waves are steady states in a co-rotating system in the complex plane, i.e.,  we replace $\partial_t \mapsto \partial_t - \mathrm{i} \Omega_0$ and restrict the global rotation in the complex plane with a phase condition.

\section{Turing and wave bifurcations}\label{app:Turing-wave}
We perform a local bifurcation analysis around the trivial state (Null phase) of the NRSH model~\eqref{eq:NRSH} and the one-mode approximation~\eqref{eq:nr_AE}, which is derived by focusing on dominant spatial Fourier modes.

\subsection{Destabilization from Null phase}
First, we linearize Eq.~\eqref{eq:NRSH} around the Null state and transform it into wavenumber space.
The dispersion relation (eigenvalues) of the linear matrix is obtained as follows:
\begin{equation}
    \lambda_\pm(k) = \varepsilon - (k_{\mathrm{c}}^2 - k^2)^2 \pm \sqrt{\chi^2 + \delta^2 - \alpha^2}.
\end{equation}
The mode $|k| = k_{\mathrm{c}}$ is most susceptible to instability.
The eigenvalue with the largest real part is given by $\lambda_+(k = k_{\mathrm{c}}) = \varepsilon + \sqrt{\chi^2 + \delta^2 - \alpha^2}$.
Regarding the destabilization of the Null phase as $\varepsilon$ increases, the sign of $\chi^2 + \delta^2 - \alpha^2$ in the root determines whether the eigenvalue at the time of destabilization has an imaginary component.
When $\chi^2 + \delta^2 - \alpha^2 > 0$, a real eigenvalue becomes unstable, indicating a Turing bifurcation:
\begin{equation}
    \varepsilon + \sqrt{\chi^2 + \delta^2 - \alpha^2} = 0.
    \label{eq:Turing-bifurcation}
\end{equation}
Note that Eq.~\eqref{eq:Turing-bifurcation} in the one-mode approximation \eqref{eq:nr_AE} corresponds to a pitchfork bifurcation.
This suggests that the instability is driven by mechanisms similar to passive interactions, representing a static phase transition.

On the other hand, when the nonreciprocity $\alpha$ is relatively strong, i.e., when $\chi^2 + \delta^2 < \alpha^2$, a complex eigenvalue with nonzero imaginary part becomes unstable, which indicates a wave bifurcation.
This essentially means that the system is active and is a type of dynamical phase transition caused by nonreciprocity $\alpha$.
The wave bifurcation occurs when
\begin{equation}
    \varepsilon = 0.
    \label{eq:wave-bifurcation}
\end{equation}

Next, we compute the normal form of the Turing bifurcation (O(2)-symmetric pitchfork) and wave bifurcation (O(2)-symmetric Hopf) using standard multiscale analysis.
The one-mode approximation \eqref{eq:nr_AE} for $\vec \phi_1 = (\phi_1,\psi_1)$ is given by
\begin{equation}\label{eq:lin_op}
\begin{split}
    \dot{\vec \phi}_1 = \tensg{J} \vec \phi_1 + \vec {N}(\vec \phi_1) \quad \text{with} \quad
    \tensg{J} &= \begin{bmatrix}
        \varepsilon + \delta & -(\chi + \alpha)
        \\
        -(\chi - \alpha) & \varepsilon - \delta
    \end{bmatrix}\\
    \text{and}\qquad
    \vec{N}(\vec{\phi_1})&=\begin{bmatrix}
        -3\phi_1 |\phi_1|^2\\
        -3\psi_1 |\psi_1|^2
    \end{bmatrix}.
\end{split}
\end{equation}

\subsection{Turing bifurcations}
Close to the Turing bifurcation, $\tensg J$ has a real eigenvalue $|\lambda| \coloneqq \sigma^2 \ll 1$, i.e., the bifurcation condition in Eq.~\eqref{eq:Turing-bifurcation} holds to leading order, and the linearized system at leading order has an eigenvector $\vec{e}$ and adjoint eigenvector $\vec{e}^\dagger$ with zero eigenvalue given by
\begin{equation}
    \vec{e} = \frac{1}{N} \begin{bmatrix}
        -(\chi+\alpha)
        \\
        \varepsilon+\delta
    \end{bmatrix},\qquad
    \vec{e}^\dagger = \frac{-\mathrm{sgn}(\varepsilon)}{N} \begin{bmatrix}
        \varepsilon-\delta
        \\
        -(\chi+\alpha)
    \end{bmatrix}
\end{equation}
where $N = \sqrt{|2\varepsilon(\chi+\alpha)|}$ is a normalization factor, chosen for orthonormality with the adjoint system, i.e., $\langle \vec e^\dagger ; \vec e \rangle = 1$.
We use the ansatz $\vec\phi_1 = \sigma Z(T)\vec v + \mathrm{h.o.t.}$ with complex amplitude $Z(T)$ and a slow timescale $T = \sigma^2 t$.
Inserting into the nonlinearity $\vec{N}(\vec{\phi_1})$ yields

\begin{align}
    \vec{N}(\vec{\phi_1})
    =-\sigma^3 Z|Z|^2 \frac{3}{N^3}
    \begin{bmatrix}
       -(\chi+\alpha)^3\\
       (\varepsilon+\delta)^3
   \end{bmatrix}
   +\mathrm{h.o.t.}
\end{align}

After projecting the equation at order $\sigma^3$ onto $\vec{e}^\dagger$, i.e., applying a Fredholm alternative, we obtain the leading order amplitude equation
\begin{subequations}
    \begin{align}
        \partial_T Z &= \pm Z - k Z|Z|^2 + \mathrm{h.o.t.}
        \\
         \text{with} \quad k &= \frac{3 \mathrm{sgn}(\varepsilon)}{4 \varepsilon^2 (\chi + \alpha)}\left[(\chi + \alpha)^2 (\varepsilon - \delta) + (\varepsilon + \delta)^3\right].\label{eq:third_order_coeff}
    \end{align}
\end{subequations}

In particular, the Turing bifurcation is supercritical if $\varepsilon[(\chi + \alpha)^2 (\varepsilon - \delta) + (\varepsilon + \delta)^3] > 0$ and the codimension-two point where the cubic coefficient is degenerate is given by:
\begin{subequations}
    \label{eq:Turing_tricricital}
    \begin{align}
        (\chi + \alpha)^2 (\varepsilon - \delta) + (\varepsilon + \delta)^3 &= 0,
        \\
        \delta^2 - \varepsilon^2 + \chi^2 - \alpha^2 &= 0.
        \label{eq:Turing_manifold}
    \end{align}
\end{subequations}

Subcriticality of the primary Turing bifurcation ($\varepsilon < 0$) occurs only for $\delta < \varepsilon < 0$ and $\alpha > \chi$ within our sign convention ($\chi > 0$ and $\alpha > 0$).
To obtain this result, one may, e.g., analyze the sign of $k$ in Eq.~\eqref{eq:third_order_coeff} in different segments of the Turing bifurcation manifold \eqref{eq:Turing_manifold}, which is circular in the coordinates $(\varepsilon, \alpha)$.
For $\delta < 0$ the relevant segment starts at $(\varepsilon = \delta, \alpha = \chi)$, where $k > 0$ and ends at $(\varepsilon \lessapprox 0, \alpha \lessapprox \sqrt{\chi^2 + \delta^2})$, where $k < 0$, which implies a change from super- to subcriticality in this segment.

\subsection{Wave bifurcations}
Close to a wave bifurcation, $\tensg{J}$ has two complex conjugate eigenvalues with frequency $\omega^2=\alpha^2-\chi^2-\delta^2>0$ and small real part $|\varepsilon|\coloneqq\sigma^2\ll 1$. At onset the eigenvectors $\vec{e}_{\mathrm{L, R}}$ and the corresponding adjoint eigenvectors $\vec e_{\mathrm{L, R}}^\dagger$ are

\begin{equation}
    \vec e_{\mathrm{L, R}} = \frac{1}{N_{\mathrm{L, R}}}\begin{bmatrix}
    \alpha+\chi
    \\
    \delta\mp\mathrm{i}\omega
    \end{bmatrix}\qquad \vec e_{\mathrm{L, R}}^\dagger=\frac{1}{N_{\mathrm{L, R}}^\dagger}\begin{bmatrix}
    \alpha - \chi
    \\
    -\delta\mp\mathrm{i}\omega
    \end{bmatrix}
\end{equation}
with orthonormality $\langle \vec e_L^\dagger; \vec e_R \rangle = 0$, $\langle \vec e_{\mathrm{L, R}}^\dagger; \vec e_{\mathrm{L, R}}\rangle = 1$, $|N_L|^2 = |N_R|^2 = 2 \alpha (\alpha + \chi)$ and $N_{\mathrm{L, R}} N_{\mathrm{L, R}}^\dagger = 2 \omega (\omega \mp \mathrm{i} \delta)$.
The ansatz at leading order is $\vec \phi_1 = \sigma(Z_{\mathrm{L}}(T) \vec e_L e^{\mathrm{i}\omega t} + Z_{\mathrm{R}}(T) \vec e_R e^{-\mathrm{i} \omega t}) + \mathrm{h.o.t.}$, where $Z_{\mathrm{L}}(T)$ and $Z_{\mathrm{R}}(T)$ represent complex amplitudes of left- and right-traveling waves (anti-clockwise and clockwise rotating amplitudes in the complex plane).
At order $\sigma^3$, we obtain the leading order equations via projection onto $\vec e_{\mathrm{L, R}}^\dagger e^{\pm\mathrm{i}\omega t}$\begin{equation}\label{eq:normal_form_wave}
\begin{split}
    \partial_T \begin{bmatrix}
       Z_{\mathrm{L}}
       \\
       Z_{\mathrm{R}}
   \end{bmatrix}=&\pm  \begin{bmatrix}
       Z_{\mathrm{L}}
       \\
       Z_{\mathrm{R}}
   \end{bmatrix}+\frac{3}{2}k_1(|Z_{\mathrm{L}}|^2+|Z_{\mathrm{R}}|^2)\begin{bmatrix}
       Z_{\mathrm{L}}
       \\
       Z_{\mathrm{R}}
   \end{bmatrix}\\&+\frac{k_1}{2}(|Z_{\mathrm{R}}|^2-|Z_{\mathrm{L}}|^2)\begin{bmatrix}
       Z_{\mathrm{L}}
       \\
       -Z_{\mathrm{R}}
   \end{bmatrix}.
\end{split}
\end{equation}
The coefficient $k_1$ is given by
\begin{equation}
    k_1 = -3 \left(1 + \mathrm{i} \frac{\delta\chi}{\alpha\omega}\right).
\end{equation}
Equations~\eqref{eq:normal_form_wave} are the (Poincaré-Birkhoff) normal form of the O(2)-symmetric Hopf bifurcation, e.g., discussed by Crawford and Knobloch \cite{CrKn1988pd}, given by
(cf.\ Eq.~1.4 of Ref.~\onlinecite{CrKn1988pd})
\begin{equation}\label{eq:pb_normal_form_wave}
\begin{split}
    \partial_T \begin{bmatrix}
       Z_{\mathrm{L}}\\
       Z_{\mathrm{R}}
   \end{bmatrix}=(p+\mathrm{i}q)\begin{bmatrix}
       Z_{\mathrm{L}}\\
       Z_{\mathrm{R}}
   \end{bmatrix}+(r+\mathrm{i}s)(|Z_{\mathrm{R}}|^2-|Z_{\mathrm{L}}|^2)\begin{bmatrix}
       Z_{\mathrm{L}}\\
       -Z_{\mathrm{R}}
   \end{bmatrix},
\end{split}
\end{equation}
where $p$, $q$, $r$ and $s$ may depend on specific combinations of $Z_{\mathrm{L}}$ and $Z_{\mathrm{R}}$.
Matching the coefficients yields (to leading order)
\begin{subequations}
    \begin{align}
        p&=\pm 1+\frac{3}{2}\Re{k_1}=\pm 1 -\frac{9}{2}(|Z_{\mathrm{L}}|^2+|Z_{\mathrm{R}}|^2) \\
        q&=\frac{3}{2}\Im{k_1}=-\frac{3\delta \chi}{2\alpha \omega}(|Z_{\mathrm{L}}|^2+|Z_{\mathrm{R}}|^2)\\
        r&=\frac{1}{2}\Re{k_1}=-\frac{3}{2}\\
        s&=\frac{1}{2}\Im{k_1}=-\frac{1\delta \chi}{2\alpha \omega}
    \end{align}
\end{subequations}
The parameters that determine stability in the non-degenerate normal form (table I, case I of Ref.~\onlinecite{CrKn1988pd}) only depend on the real coefficients $p$ and $r$ and are given by $\mathrm{sgn}(r)=-1$ and $\mathrm{sgn}(r) p_N/r=-3$.
Hence, there is only one possible bifurcation scenario in our system (Upper case in Fig.~1 of Ref.~\onlinecite{CrKn1988pd} with inverted stability assignments since $\mathrm{sgn}(r)=-1$). In particular, traveling and standing waves will always emerge supercritical with traveling waves being stable and standing waves unstable.

\section{Bifurcations in the nonlinear regime}\label{app:local-bif-derivation}
In the following, we focus on the one-mode approximation~\eqref{eq:nr_AE}, which retains only the complex amplitudes of the main spatial Fourier modes.
As discussed in the main text, it can be further reduced to the amplitude-phase representation \eqref{eq:phase_difference} due to the underlying O(2)-symmetry.
In this representation, the static phase (SS) and the dynamical phase (TW), which emerge from the Null state at its instability threshold, correspond to fixed points.
We use this to investigate secondary bifurcations, such as the destabilization and annihilation of these phases.
If we classify the fixed points based on the condition $\dot{\eta} = 0$, there are patterns where $\sin \eta = 0$ and patterns where the other factor is $0$.
The former [latter] corresponds to the fixed point equivalent of the SS phase [TW phase].
The fixed point equivalent to the SS phase is also a fixed point in the system of Eq.~\eqref{eq:nr_AE}, but the fixed point equivalent to TW in \eqref{eq:phase_difference} is not a fixed point in the system of Eq.~\eqref{eq:nr_AE}, but is mapped to a limit cycle solution with constant phase velocity.

\subsection{Fixed point (SS)}
To analyze the fixed points corresponding to the SS phase, we define $\xi \coloneqq (\varepsilon + \delta - 3 \rho_\phi^2) / (\chi + \alpha)$, and derive the fourth-order algebraic equation for $\xi$:
\begin{equation}
    0 = (\chi + \alpha) \xi^4 - (\varepsilon + \delta) \xi^3 + (\varepsilon - \delta) \xi - (\chi - \alpha)
    \label{eq:FP_SS}
\end{equation}
Each real solution of this algebraic equation corresponds to such a fixed point as long as $\xi$ satisfies the condition $3 \rho_\phi^2 = \varepsilon + \delta - (\chi + \alpha) \xi > 0$.
The parameter set for which the number of real solutions changes can be obtained from the discriminant $S$ of a quartic equation in $\xi$, giving pair creation or annihilation at the fixed point, i.e., saddle-node bifurcation, where
\begin{align}
    & 64 \alpha^{6} - 192 \alpha^{4} \chi^{2} - 48 \alpha^{4} \left(\delta - \varepsilon\right) \left(\delta + \varepsilon\right) + 192 \alpha^{2} \chi^{4}
    \notag
    \\
    & + 96 \alpha^{2} \chi^{2} \left(\delta - \varepsilon\right) \left(\delta + \varepsilon\right) - 3 \alpha^{2} \left(\delta^{2} + 5 \varepsilon^{2}\right) \left(5 \delta^{2} + \varepsilon^{2}\right)
    \notag
    \\
    & + 108 \alpha \chi \delta \varepsilon \left(\delta^{2} + \varepsilon^{2}\right)
    - 64 \chi^{6} - 48 \chi^{4} \left(\delta - \varepsilon\right) \left(\delta + \varepsilon\right)
    \notag
    \\
    & - 12 \chi^{2} \left(\delta^{4} + 7 \delta^{2} \varepsilon^{2} + \varepsilon^{4}\right) - \left(\delta - \varepsilon\right)^{3} \left(\delta + \varepsilon\right)^{3} = 0
    \label{eq:saddle-node-bifurcation}
\end{align}

\subsection{Fixed point (TW)}
We now consider the fixed point in \eqref{eq:phase_difference} characterized by $\sin\eta \ne 0$.
Under this condition, the analytical expressions for the amplitudes and the phase difference are given by:
\begin{align}
    \rho_\phi &= \sqrt{\frac{\varepsilon}{3 \alpha} (\alpha + \chi)},
    \\
    \rho_\psi &= \sqrt{\frac{\varepsilon}{3 \alpha} (\alpha - \chi)},
    \\
    \cos \eta &= - \frac{\chi}{\alpha} \frac{1}{\sqrt{\alpha^2 - \chi^2}} \left(\varepsilon - \frac{\alpha}{\chi}\delta \right).
\end{align}
This solution corresponds to a traveling wave (TW) where the amplitude remains constant while the phase progresses at a constant velocity $\Omega_0$.
Note that a special case is given by $\chi \varepsilon = \alpha \delta$, where $\cos \eta = 0$.
Substituting these fixed point values into the phase evolution equations yields the square of the phase velocity:
\begin{equation}
    \Omega_0^2 = \frac{1}{\alpha^2} \left[ \alpha^2 (\alpha^2 - \chi^2) - \chi^2 \left(\varepsilon - \frac{\alpha}{\chi} \delta\right)^2 \right].
\end{equation}

The existence of the TW solution requires $\Omega_0^2 > 0$ (which ensures $|\cos \eta| < 1$).
The limit $\Omega_0 \to 0$ marks the ceasing of motion, where the TW phase merges with the SS phase.
This transition corresponds to the drift-pitchfork bifurcation, defined by the condition:
\begin{equation}
    \alpha^2 (\alpha^2 - \chi^2) - \chi^2 \left(\varepsilon - \frac{\alpha}{\chi} \delta\right)^2 = 0.
    \label{eq:drift-pitchfork-bifurcation}
\end{equation}

Furthermore, by performing a linear stability analysis of the TW state in Eq.~\eqref{eq:phase_difference} and applying the Routh-Hurwitz criterion~\cite{gantmacher1959theory} to the characteristic polynomial of the resulting Jacobian, we obtain the condition for the instability threshold of the TW.
This corresponds to a Hopf bifurcation in the co-moving frame \eqref{eq:phase_difference} (manifesting as a Neimark-Sacker bifurcation in the stationary frame \eqref{eq:nr_AE}), defined by:
\begin{equation}
    \alpha^{4} - \alpha^{2} \chi^{2} + \alpha^{2} \left(- \delta^{2} + 2 \varepsilon^{2}\right) + 4 \alpha \chi \delta \varepsilon - 5 \chi^{2} \varepsilon^{2} = 0.
    \label{eq:Hopf-bifurcation}
\end{equation}
These analytical bifurcation conditions derived from the one-mode approximation are summarized in Table~\ref{tab:bifurcation-conditions}.

\subsection{Codimension-two bifurcation conditions}
The conditions for the codimension-two bifurcations in Table~\ref{tab:codim2_bifurcation-conditions} are obtained by requiring two codimension-one conditions to be simultaneously satisfied.
In particular:
\begin{subequations}
\begin{align}
\text{Takens-Bogdanov:}\quad &\text{Eq.~}\eqref{eq:Turing-bifurcation} \quad\land\quad \text{Eq.~}\eqref{eq:wave-bifurcation}   \\
\text{DP + SNIPer:}\quad &\text{Eq.~}\eqref{eq:drift-pitchfork-bifurcation}\quad \land \quad \text{Eq.~}\eqref{eq:Hopf-bifurcation}\label{eq:DP+Sniper}.
\end{align}
\end{subequations}
Equations~\eqref{eq:DP+Sniper} also contain the decoupled transition ${(\varepsilon = \delta > 0, \alpha = \chi)}$ as a trivial solution. Further note that one of the two conditions may be replaced with the condition \eqref{eq:saddle-node-bifurcation} for the saddle-node bifurcation, since, as we discuss in Section \ref{sec:param_space}, Hopf, saddle-node, and drift-pitchfork bifurcation generically appear simultaneously in a codimension-two point. The double SNIPer bifurcation is obtained by imposing two simultaneous saddle-node bifurcations, i.e., demanding that Eq.~\eqref{eq:FP_SS} factorizes as
\begin{equation}
    (\chi + \alpha) (\xi - a)^2 (\xi - b)^2 = 0.
\end{equation}
Comparing the coefficients of the terms $1$, $\xi$, $\xi^2$ and $\xi^3$ yields four equations that constrain $(a,b,\varepsilon,\delta,\chi,\alpha)$, where we eliminate $a$ and $b$ and solve for $\varepsilon$ and $\alpha$.

The tricritical Turing bifurcation is determined by the degeneracy of the cubic normal form coefficient, i.e., by Eq.~\eqref{eq:Turing_tricricital} in Appendix~\ref{app:Turing-wave}.

\section{Derivation of the Normal Form coefficients}\label{app:nit}
The one-mode approximation of the NRSH model~\eqref{eq:nr_AE} exhibits a Takens-Bogdanov (TB) bifurcation with O(2)-symmetry of the Null solution $(\phi_1, \psi_1) = (0, 0)$ at parameters $(\varepsilon, \alpha) = (0, \sqrt{\chi^2 + \delta^2})$.
We derive the normal form coefficients through a linear coordinate transform followed by a near-identity transformation.

At the TB bifurcation, the matrix $\tensg{J}$ in Eq.~\eqref{eq:lin_op} has a double zero eigenvalue with a geometric multiplicity of one, i.e., $\tensg{J}$ is non-diagonalizable at the TB point.
According to normal form theory, it is possible to eliminate non-resonant terms near a bifurcation point via appropriate variable transformations (combinations of linear and weakly nonlinear ones), and the resulting system constitutes a normal form for this bifurcation.
The normal form of the TB bifurcation with O(2)-symmetry has been studied in detail by Ref.~\onlinecite{DaKn1987ptrslsapes}.
We aim to leverage the comprehensive classification established by Ref.~\onlinecite{DaKn1987ptrslsapes} by reducing Eq.~\eqref{eq:nr_AE} to \eqref{eq:TB-normal} via appropriate variable transformations.
Therefore, we first need to transform the linear part to that of Eq.~\eqref{eq:TB-normal}.
Since the trace and determinant are invariant under a similarity transformation, we establish the following relation for the unfolding parameters:
\begin{equation}
    \mu = \delta^2- \varepsilon^2 + \chi^2- \alpha^2,
    \quad
    \nu = 2 \varepsilon.
    \tag{\ref{eq:mu-nu}}
\end{equation}
In particular, such a linear part can be realized by the following matrix similarity transformation:
\begin{equation}
    \tensg{P} = \begin{bmatrix}
        1 & 1
        \\
        \varepsilon + \delta - \chi - \alpha
        & \varepsilon - \delta - \chi + \alpha
    \end{bmatrix}.
    \label{eq:linear-P}
\end{equation}
Note that when $\alpha = \delta$, this linear transformation degenerates and becomes singular, but this does not occur in the parameter domain considered in the main text.

Substituting the linear transformation $[z, w]^\top = \tensg{P} [\phi_1, \psi_1]^\top$ into Eq.~\eqref{eq:nr_AE}, we obtain the general form for the set of real coefficients $(a_i, b_i, c_i)$ depending on the physical parameters $(\varepsilon, \delta, \chi, \alpha)$:
\begin{subequations}
    \begin{align}
        &\begin{aligned}
            \dot{z} ={}&
            w
            + \left(a_1 |z|^2 + b_1 |w|^2\right) z + c_1 z^2 \overline{w}
            \\
            &+ \left(
                a_2 |z|^2 + b_2 |w|^2
            \right) w
            + c_2 w^2 \overline{z},
            \end{aligned}
            \\
            &\begin{aligned}
            \dot{w} ={}&
            \mu z + \nu w
            + \left(a_3 |z|^2 + b_3 |w|^2\right) z
            + c_3 z^2 \overline{w}
            \\
            &+ \left(a_4 |z|^2 + b_4 |w|^2\right) w
            + c_4 w^2 \overline{z},
        \end{aligned}
        \label{eq:TB-general}
    \end{align}
\end{subequations}
The linear transformation introduces apparent nonlinear couplings through the original third-order nonlinear terms. However, as discussed in Section~\ref{sec:param_space}, this system is inherently only weakly coupled via linear terms, resulting in decoupling in the $\alpha \to \chi$ limit.

We use a third-order near-identity transformation preserving O(2)-symmetry to eliminate non-resonant nonlinear terms.
This reduces Eq.~\eqref{eq:TB-general} to the normal form of a TB bifurcation with O(2)-symmetry in Eq.~\eqref{eq:TB-normal}.

By performing a near-identity transformation
\begin{subequations}
    \begin{align}
        &\begin{aligned}
            v ={}&
            v' + \left(\alpha_1 |v|^2 + \beta_1 |w|^2\right) v + \gamma_1 v^2 \overline{w}
            \\
            &+ \left(\alpha_2 |v|^2 + \beta_2 |w|^2\right) w + \gamma_2 w^2 \overline{v},
        \end{aligned}
        \\
        &\begin{aligned}
            w ={}&
            w' + \left(\alpha_3 |v|^2 + \beta_3 |w|^2\right) v + \gamma_3 v^2 \overline{w}
            \\
            &+ \left(\alpha_4 |v|^2 + \beta_4 |w|^2\right) w + \gamma_4 w^2 \overline{v}
        \end{aligned}
    \end{align}
\end{subequations}
with appropriate parameter combinations:
\begin{subequations}
    \begin{align}
        \alpha_{1} &= - \frac{\mu b_{2}}{2} + \frac{\nu^{2} b_{2}}{3} - \frac{\nu \left(b_{1} + b_{4}\right)}{2} + \frac{a_{2} + c_{4}}{2},
        \\
        \alpha_{2} &= - \frac{4 \nu b_{2}}{3} + \frac{b_{1} + b_{4} + c_{2}}{2},
        \\
        \alpha_{3} &= - \mu \nu b_{2} + \frac{\mu \left(b_{1} + b_{4} + c_{2}\right)}{2} - a_{1},
        \\
        \alpha_{4} &= \mu b_{2} - \frac{2 \nu^{2} b_{2}}{3} - \frac{\nu \left(b_{1} + b_{4} - c_{2}\right)}{2} + c_{4},
        \\
        \beta_{1} &= 0,
        \\
        \beta_{2} &= 0,
        \\
        \beta_{3} &= - \frac{2 \nu b_{2}}{3} - \frac{b_{1} - b_{4} - c_{2}}{2},
        \\
        \beta_{4} &= 0,
        \\
        \gamma_{1} &= \frac{\nu b_{2}}{3},
        \\
        \gamma_{2} &= b_{2},
        \\
        \gamma_{3} &= - \frac{\mu b_{2}}{2} + \frac{2 \nu^{2} b_{2}}{3} - \frac{\nu \left(b_{1} + b_{4}\right)}{2} + \frac{a_{2} - 2 c_{1} + c_{4}}{2},
        \\
        \gamma_{4} &= \frac{2 \nu b_{2}}{3} + \frac{b_{1} + b_{4} - c_{2}}{2},
    \end{align}
\end{subequations}
we obtain the following correspondence between Eq.~\eqref{eq:TB-normal} and the physical parameters:
\begin{subequations}
    \begin{align}
        A &= - \mu^{2} b_{2} + \mu \nu \left(b_{1} + b_{4}\right) + \mu \left(c_{1} - c_{4}\right) - \nu a_{1} + a_{3},
        \\
        B &= \nu \left(2 b_{1} + b_{4} - c_{2}\right) - a_{2} + b_{3} + 2 c_{1} - 2 c_{4},
        \\
        C &= 2 \mu \nu b_{2} + \mu \left(- b_{4} - c_{2}\right) + a_{1} +c_{3},
        \\
        D &= - 2 \mu \nu b_{2} + \mu \left(- b_{1} - b_{4} + c_{2}\right) + a_{1} + a_{4} - c_{3}.
    \end{align}
    \label{eq:A~D_mu_nu}
\end{subequations}
Note that these expressions neglect terms of fifth order of $z$ and $w$ and third order of $\mu$ and $\nu$ through the near-identity transformation.

Finally, by combining the coefficients $(a_i, b_i, c_i)$ obtained from Eq.~\eqref{eq:TB-general} with the relations in Eq.~\eqref{eq:A~D_mu_nu}, we determine the explicit dependence of the normal form parameters $A$ through $D$ (and $M$, where $M = 2 C + D$) on the physical parameters, as listed in Eq.~\eqref{eq:A~M-NRSH} in the main text.

\section{Global bifurcation and dynamical bifurcation conditions from the normal form of Takens-Bogdanov bifurcation with O(2)-symmetry}\label{app:global-bif-results}
The global and dynamical bifurcation conditions for the normal form of the Takens-Bogdanov bifurcation with O(2)-symmetry were derived by Dangelmayr and Knobloch~\cite{DaKn1987ptrslsapes}.
These results, expressed in terms of the real unfolding parameters $\mu$ and $\nu$, are valid in the immediate vicinity of the codimension-two point.
While the conditions in Table~\ref{tab:bifurcation-conditions} are obtained through a full stability analysis of the fixed points within the one-mode approximation and remain valid globally in the $(\varepsilon, \alpha)$-plane, the expressions presented here represent leading-order asymptotic approximations valid only in the vicinity of the Takens-Bogdanov point.

The codimension-one bifurcation manifolds emerging from the Takens-Bogdanov point, as illustrated in Fig.~\ref{fig:tb_quantitative}, are given by the following relations among the unfolding parameters $\mu$ and $\nu$ and the normal form coefficients $A$, $D$, and $M$.
The drift-pitchfork (DP) bifurcation is approximated by
\begin{equation}\label{eq:TB_DP}
    A \nu - D \mu = 0.
\end{equation}
The Hopf (H) bifurcation condition for the TW state is expressed as\footnote{
The general condition is $\mu = [(3M - 5D) / (2 M - 4 D)] A \nu / D$, but substituting the value $D / M = 1 / 3$ for our model simplifies this to the form shown here.
}
\begin{equation}\label{eq:TB_H}
    2 A \nu - D \mu = 0.
\end{equation}
The heteroclinic (Het) bifurcation, where the standing wave orbit collides with the steady-state fixed points, is given by
\begin{equation}\label{eq:TB_Het}
    5 A \nu - 4 M \mu = 0.
\end{equation}
The drift-pitchfork bifurcation of limit cycles (DPLC), which relates the modulated waves and standing waves, is expressed as
\begin{equation}\label{eq:TB_DPLC}
    A \nu - \frac{2 D (1 - \Phi(k))}{1 + k^2} \mu = 0.
\end{equation}
Here, $\Phi(k)$ is defined as the ratio of the complete elliptic integrals of the second kind to those of the first kind, $E(k) / K(k)$.
The modulus $k$ is determined by the ratios of the nonlinear coefficients. For the internal algebraic constraint $D / M = 1 / 3$ found in this model, the modulus satisfies the following:
\begin{equation}
    \frac{1}{3} = \frac{(1 - k^2) (2 - k^2) - 2 (1 - k^2 + k^4) \Phi(k)}{5 (1 - \Phi(k)) [1-k^2 - (1+k^2) \Phi(k)]}.
\end{equation}
Numerical evaluation yields $k \approx 0.9415$ and $\Phi(k) \approx 0.3772$.

%


\end{document}